\documentclass[prb,aps,twocolumn,floats,showpacs]{revtex4}
\usepackage{epsfig}
\usepackage{graphicx}

\newcommand{\bfig}{\begin{figure}}
\newcommand{\efig}{\end{figure}}
\newcommand{\be}{\begin{equation}}
\newcommand{\ee}{\end{equation}}
\newcommand{\bea}{\begin{eqnarray}}
\newcommand{\eea}{\end{eqnarray}}

\begin{document}

\title{Orbital fluctuations and strong correlations in quantum dots}

\author{Gergely Zar{\'a}nd}
\affiliation{
Theoretical Physics Department,  Institute
of Physics, Budapest Univ. of Technology and Economics, Budafoki \'ut 8, H-1521
}
\date{\today}

\begin{abstract}
In this lecture note we focus our attention to  quantum dot systems  where
exotic strongly correlated behavior develops due to the presence of orbital 
or charge degrees 
of freedom. After giving a concise overview of the theory of transport 
and Kondo effect through a single electron transistor, we discuss how 
$SU(4)$ Kondo effect develops in  dots having orbitally 
degenerate states and in double dot systems, and  
then study the singlet-triplet transition in lateral quantum dots.
Charge fluctuations and Matveev's mapping to the two-channel Kondo 
model in the vicinity of  charge degeneracy point are also discussed. 
\end{abstract}
\pacs{75.20.Hr, 71.27.+a, 72.15.Qm}
\maketitle

\section{Introduction}

Although no rigorous definition exists, we typically call a  'quantum dot' a small
artificial  structure  containing conduction electrons, and weakly  coupled to the rest  of the  
world. There is a variety of ways to produce these structures: Maybe the most common technique 
to do this is by defining a typically $\mu{\rm m}$-size region by 
shaping a two-dimensional electron gas using 
gate electrodes placed on the top of a semiconductor heterostructure or by 
etching (see, {\em e.g.}, Refs.~\onlinecite{David,TaruchaST,review}).
In Fig.~\ref{fig:david} we show the top view of such a
single electron transistor (SET) that has been first used to detect the Kondo effect
in such a structure.\cite{David}  
Beside semiconductor technologies, quantum dots 
can also be built from metallic grains \cite{Devoret,Ralph_metallic}, 
and more recently it became possible to 
integrate even real molecules into electronic circuits\cite{molecular_SET}.
The common feature of all these devices is that Coulomb correlations 
play an essential role in them, and induce  Coulomb blockade\cite{Coulomb_blockade}
 and  Kondo  effect.\cite{David,Leo,Kondo_dot}

\bfig
\includegraphics[width=6cm]{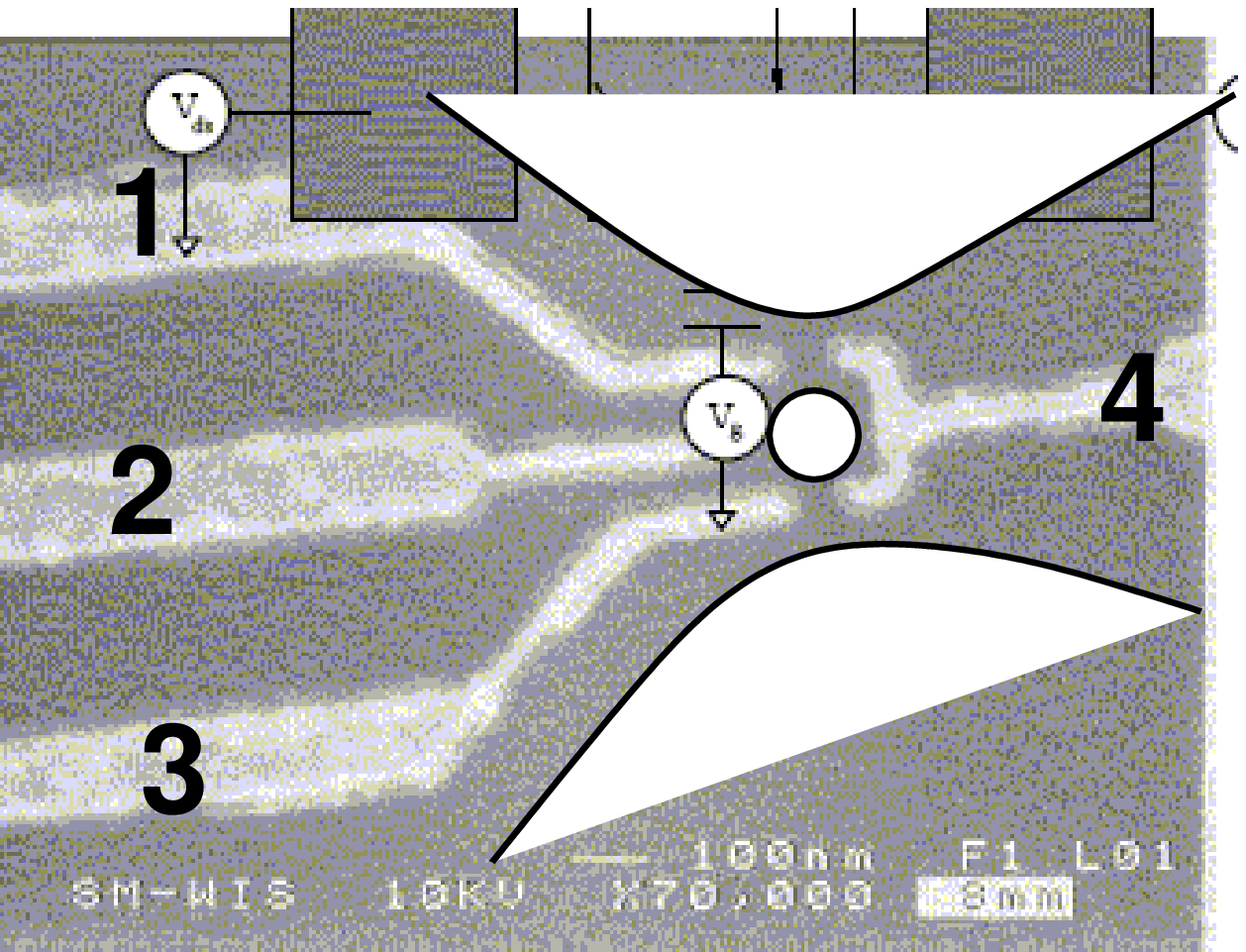}
\caption{
\label{fig:david}
Top view of the SET used by David Goldhaber-Gordon and his collaborators 
to first observe the Kondo effect in a quantum dot, from Ref.~\onlinecite{David}: 
The white areas indicate regions where conduction electrons are present. 
The quantum dot is at the central region (white circle). The various electrodes 
(1-4) have been used to define the dot and the junctions.
}
\efig

In the present paper (which has been prepared as a lecture note) we
 shall not attempt 
to give a complete overview of this enormous field. Instead, we shall  
 first give  a concise  introduction into   the basic properties of these 
devices,  and then focus our attention to some exotic strongly correlated states 
associated with orbital and charge degrees of freedom that appear in them. 

There is two essential energy scales that  characterize an isolated quantum dot: 
One of them is the {\em charging energy}, $E_C$,  the typical cost of putting  
an extra electron on the device. The other is the typical separation of single 
particle energies, also called level spacing,  $\Delta$. Typically $\Delta \ll E_C$, but for 
very small structures ({\em e.g.} in the extreme case of a molecule) these two energy 
scales can be of the same order of  magnitude. 
While the charging energy $E_C$ is usually of the order of 
$e^2/L$ with $L$ the characteristic 
size of the device, the level spacing  $\Delta$ depends very much 
on the  material and dimensionality of the dot: it is typically 
very small in mesoscopic metallic grains, where it roughly scales 
as $\Delta \sim E_F/(k_F L)^3$, and becomes of the order of one Kelvin  only for 
nanoscale structures with $L \sim 20 \AA$. 
For two-dimensional semiconductor structures, on the other hand,
both $k_F$ and the Fermi energy are much smaller and  since $\Delta$ scales as  
$\Delta\sim E_F/(k_F L)^2$, in these structures 
$\Delta$ becomes of the order of a Kelvin typically for $L\sim 0.1 \mu{\rm m}$.

Coulomb correlations may become  only important if the measurement temperature is less than 
the charging energy, $T<E_C$. Clearly, this criterion can be only satisfied 
with our current cooling technology if $E_C$ is in the range of a few Kelvins, 
{\em i.e.} the size of the system is in the $\mu{\rm m}$ range or below.  
The behavior of a quantum dot is also very different in the 
regimes $T > \Delta$ and $T<\Delta$: while in the former regime 
electron-hole excitations on the dot are important, for  $T<\Delta$ these
excitations do not play an essential role.  

Beside the difference in the 
typical energy scales $E_C$ and $\Delta$, there is also a difference in the 
way semiconducting and metallic devices are usually connected.
Although metallic particles can also be contacted through single or few mode 
contacts using {\em e.g.} STM tips, these grains  are typically connected through 
multichannel leads with  contact sizes much larger than the Fermi wavelength $\lambda_F$.
Lateral semiconducting devices are, on the other hand, usually contacted 
through few or single mode contacts (though {\em e.g.} vertical dots 
are connected through a large contact area and thus many channels).  
While these details can be important for 
some phenomena,\cite{Schoen,Wilhelm}  the behavior of all these devices is very 
similar in  many respects. In the following, we shall therefore mainly focus on 
lateral quantum dots with single mode contacts. 

This lecture is organized as follows: First, in Section~\ref{sec:blockade}
we shall discuss the phenomenon of Coulomb blockade and the basic Hamiltonians
that are used to describe  quantum dots.
In Section~\ref{sec:Kondo} we discuss how the Coulomb blockade is lifted 
by the formation of a strongly correlated Kondo state at very low 
temperatures, $T\ll \Delta$. In Section~\ref{sec:orbital} we shall study 
more exotic strongly correlated states that appear due to 
orbital degrees of freedom, the  $SU(4)$ Kondo state, and  
the so-called singlet-triplet transition. Section~\ref{Matveev}
is devoted to the analysis of the Coulomb blockade staircase
in the vicinity of the degeneracy points, where strong charge fluctuations 
are present.

\section{Coulomb blockade}
\label{sec:blockade}

In almost all systems discussed in the introduction we can
describe the isolated dot by the following second quantized  Hamiltonian
\begin{equation}
H_{\rm dot} = \sum_{j,\sigma} \epsilon_j \;d^\dagger_{j \sigma} d_{j \sigma} 
+ H_{\rm int} + H_{\rm gate}\;,
\label{eq:dot1}
\end{equation} 
where the second  term describes interactions between the conduction electrons on the 
island, and the effects of various gate voltages are accounted for by the last term.
The operator $d^\dagger_{j \sigma}$ creates a conduction electron in a single 
particle state $\varphi_j$ with spin $\sigma$ on the dot.

Fortunately, the  terms $H_C= H_{\rm int} + H_{\rm gate}$ can be replaced 
in most cases  to a very good accuracy by a simple classical interaction term 
\cite{Aleiner}
\be  
H_C = {e^2\over 2C} \left( n_{\rm dot}  - {V_g C_g\over e}\right)^2\;,
\label{eq:H_C1}
\ee
where $C$ denotes the total capacitance of the dot, $C_g$ is the gate 
capacitance, $e$ is the electron's charge, $V_g$ stands for the gate 
voltage (roughly proportional to the voltage on electrode '2' in Fig.~\ref{fig:david}), 
and $ n_{\rm dot} = \sum_{j,\sigma} :d^\dagger_{j\sigma} d_{j\sigma}$ is the number 
of extra electrons on the dot.  This simple form can be derived by estimating the 
Coulomb integrals  for a chaotic dot \cite{Aleiner}, however, it
also follows from the phenomenon  of {\em screening} in a metallic particle. 
We outline this rather instructive derivation  of Eq.~(\ref{eq:H_C1}) within the Hartree  
approximation in Appendix~\ref{app:Hartree}. 
Clearly, the dimensionless gate voltage $N_g = {V_g C_g/ e}$
sets the number of electrons on the dot, $ \langle n_{\rm dot}\rangle  \approx N_g$.

The single particle levels $\epsilon_j$ above are random but correlated: The distribution 
of these  levels for typical ({\em i.e.} large and chaotic) islands is given with a good
accuracy by random matrix  theory,\cite{RandomMatrix} which predicts  among others 
that the separation  $s$ between two neighboring states displays a universal 
distribution, 
\be 
P(s) = {1\over \Delta} p\left({s/ \Delta}\right)\;,
\ee
with $\Delta$ the average level spacing between two neighboring levels. For small 
separations, energy levels repel each-other and $p$ vanishes as 
 $p\left({s\over \Delta}\right)\sim \left({s\over \Delta}\right)^\beta$, where the exponent 
is  $\beta=1,2$ or $4$, depending on the symmetry of the Hamiltonian
(orthogonal, unitary, and simplectic, respectively). 
In some special 
cases  cross-overs between various universality classes can also 
occur,
and in some cases level repulsion may be even absent for dots with special  symmetry 
properties.

\bfig
\includegraphics[width=5cm]{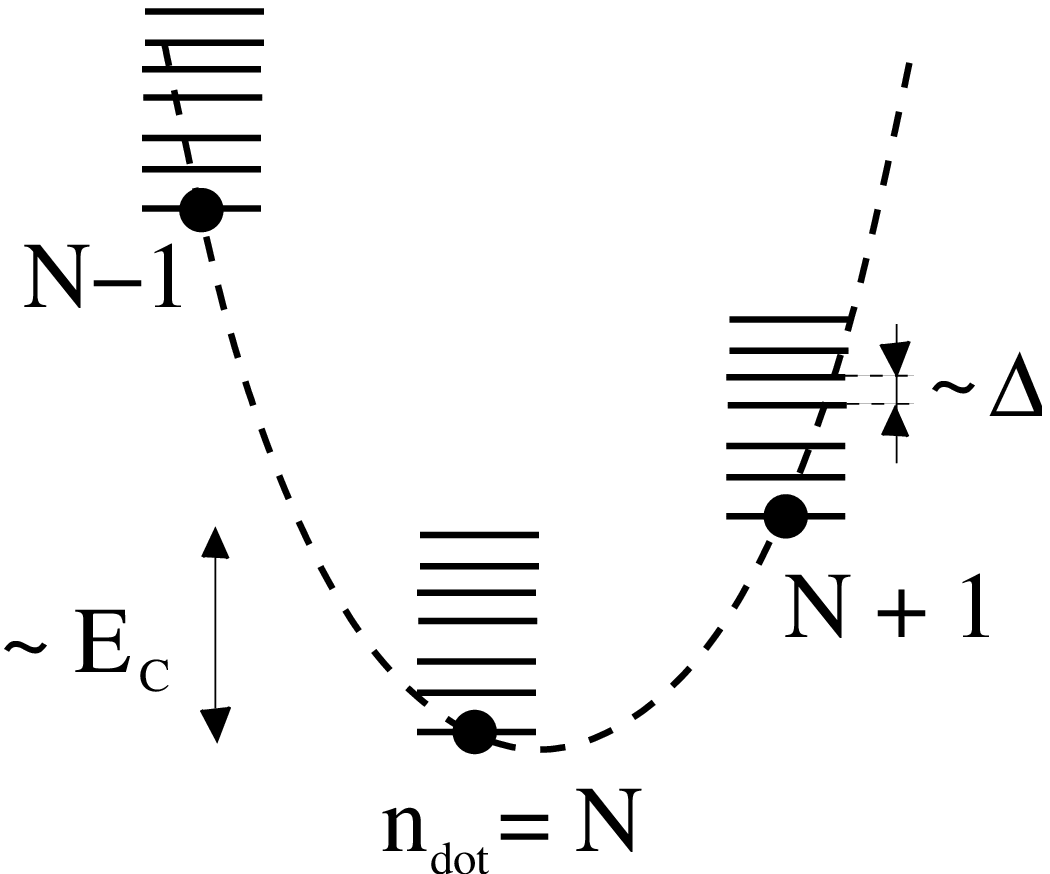}
\caption{\label{fig:spectrum} Excitation spectrum of an island. Lines represent 
eigenenergies of the island. Charging excitations typically 
need an energy $\sim E_C$ while internal electron-hole and spin excitations 
cost an energy $\Delta \ll E_C$.
}
\efig
 The spectrum of an isolated dot described by 
Eqs.~(\ref{eq:dot1}) and (\ref{eq:H_C1}) is 
sketched in Fig.~\ref{fig:spectrum}.
As we already mentioned,
for typical parameters and relatively large lateral dots or metallic islands, 
the charging energy is much larger than the  level spacing, $E_C\gg \Delta$. 
Accordingly, internal electron-hole  excitations cost much smaller energy 
than charge excitations of the dot.
For   dot sizes in the $0.1\mu {\rm m}$ range the capacitance 
$C$  can be small enough so that  the charging energy 
$E_C = e^2/2C$ associated with putting an extra electron 
on the dot can safely be in the  ${\rm meV}$ range. Therefore, 
unless $N_g = {V_g C_g\over e}$ is a half-integer, it costs 
a finite energy to charge the device, and therefore 
the number of electrons  on the dot becomes  quantized at 
low enough temperatures,  $T\ll E_C$,   and 
a Coulomb blockade develops - provided that quantum fluctuations 
induced by coupling the dot to leads are not very strong. 

Let us now  consider a quantum dot that is
weakly tunnel-coupled to leads, 'weak coupling' in this context meaning  
that  the  conductance between the island and the leads is less than the quantum
conductance, $G_Q\equiv 2e^2/h$.  In the particular case of a lateral quantum dot 
this condition is satisfied when the  last conduction electron channel is being pinched 
off. If the lead-dot conductance is much larger than $G_Q$ then the Coulomb blockade
is lifted by quantum fluctuations (see also Section~\ref{Matveev}). 
In the absence of tunneling processes, the charge on the dot would change in 
sudden steps at $T=0$ temperature (see Fig.~\ref{fig:steps}).
However, in the vicinity of the jumps, where $N_g = {V_g C_g\over e}$ is a half-integer, 
two charging states of the island become almost degenerate. Therefore
quantum tunneling to the leads induces quantum fluctuations  between  these two charging 
states, smears out the steps, and eventually
completely suppresses the steps only leaving some small oscillations on the top 
of a linear $\langle n_{\rm dot}\rangle(V_g)$ dependence.\cite{lifting_blockade}
At a finite temperature $T\ne 0$, thermal fluctuations play a similar role, 
and the charging steps are also  washed out if $T\gg E_C$.

\bfig
\includegraphics[width=8cm]{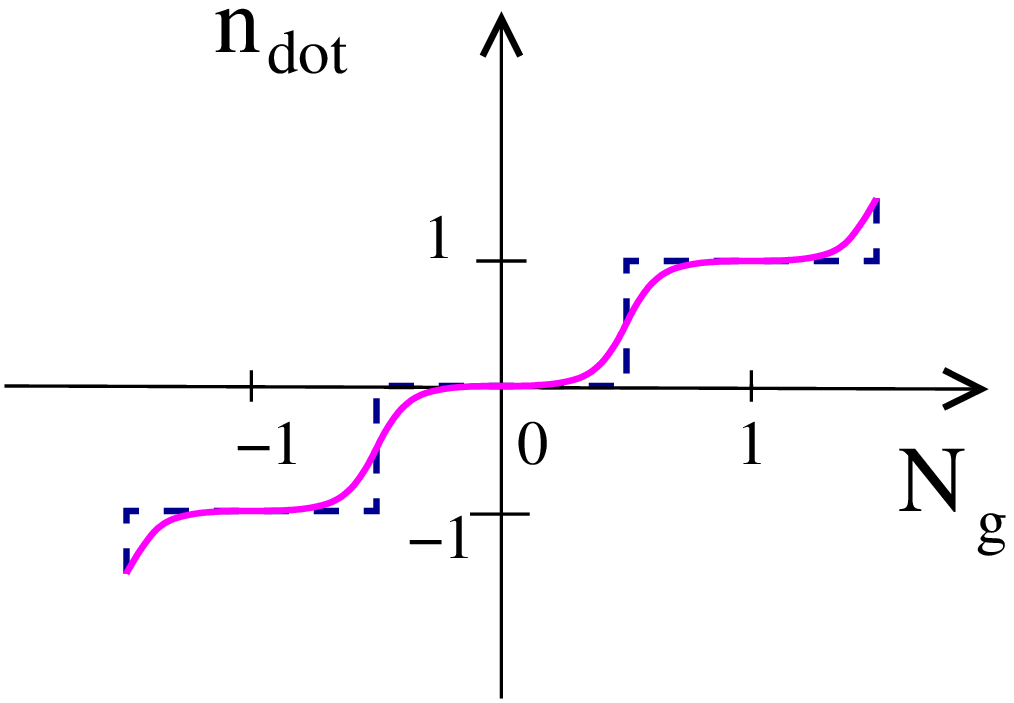}
\caption{\label{fig:steps}
The number of electrons on a quantum dot as a function of the dimensionless 
gate voltage. The sudden jumps of an isolated dot  become smeared out 
due to quantum fluctuation as soon as we couple the dot to leads.}
\efig

Charge quantization is also reflected in the transport properties of the dot.
To study transport through a quantum dot, one typically builds a single electron 
transistor (SET) by attaching the dot to two leads,
as shown in  Figs.~\ref{fig:david} and \ref{fig:SET}.
Let us first assume that the conductances $G_L$ and $G_R$ between the dot and the 
lead on the left / right are small compared to $G_Q$ and that 
 quantum fluctuations are small. 
In this limit we can describe charge fluctuations on the dot by the following simple tunneling 
Hamiltonian:
\be
\hat V = \sum_j \sum_\epsilon
\Bigl\{ t^L_j\; d^\dagger_{j \sigma } \psi_{L,\epsilon \sigma}
+ {\rm h.c.} \Bigr\}
+ ``L \leftrightarrow R'' \;,
\ee
where we assumed single mode contacts.
The fields  $\psi_{L,\epsilon \sigma }^\dagger$ and
$\psi_{R, \epsilon \sigma }^\dagger$ denote the creation operators of  conduction 
 electrons of energy $\epsilon$ 
and spin $\sigma$  in the left and right leads, respectively, and are normalized to 
 satisfy the anticommutation relations 
$\{\psi_{L/R, \epsilon \sigma }^\dagger, \psi_{L/R, \epsilon' \sigma'} \} =
 \delta_{\sigma\sigma'}\delta_{\epsilon\epsilon'}$.
Note that the tunneling matrix 
elements $t^L_j$ and $t^R_j$ fluctuate from level to level since they depend on the amplitude 
of the  wave function  $\varphi_j$ at the tunneling position. 
Using a simple Boltzmann equation  approach  one then finds that for $T<E_C$ 
the linear  conductance of the SET has peaks whenever two charge states of the dot 
are degenerate and $N_g$ is a half-integer.
In the regime, $\Delta\ll T \ll E_C$, in particular,  one finds that \cite{peak} 
\be 
G(T) \approx {1\over 2} {G_L G_R \over G_L + G_R} { \Delta E /T \over {\rm sinh} (\Delta E /T) }\;,
\label{activated}
\ee
where $\Delta E $ is the energy difference between the two charging states of the dot, 
and $G_{L/R} = (8\pi^2 e^2/h) \; \varrho_{\rm dot}\; \varrho_{L/R}\; \langle |t^{L/R}_j|^2\rangle_j$ is the 
tunnel conductance of the two junctions, with $ \varrho_{\rm dot}$ and 
$\varrho_{L/R}$ the density of states on the dot and the leads. Note that even for perfect 
charge degeneracy,
$\Delta E = 0$, 
the resistance of the SET is twice as large as the sum of the two junction resistances
due to Coulomb correlations. 
As we decrease the temperature, the conductance peaks become sharper and sharper,  
while the conductance between the peaks 
decreases exponentially, and the Coulomb blockade develops. 
This simple Boltzmann equation picture, however, breaks down at 
somewhat lower 
temperatures, where higher order processes and 
quantum fluctuations become important. These quantum fluctuations
 may even completely lift the Coulomb blockade 
and result in a perfect conductance at low temperatures as we shall 
explain in the next section. 

\bfig
\begin{center}
\includegraphics[width=5cm]{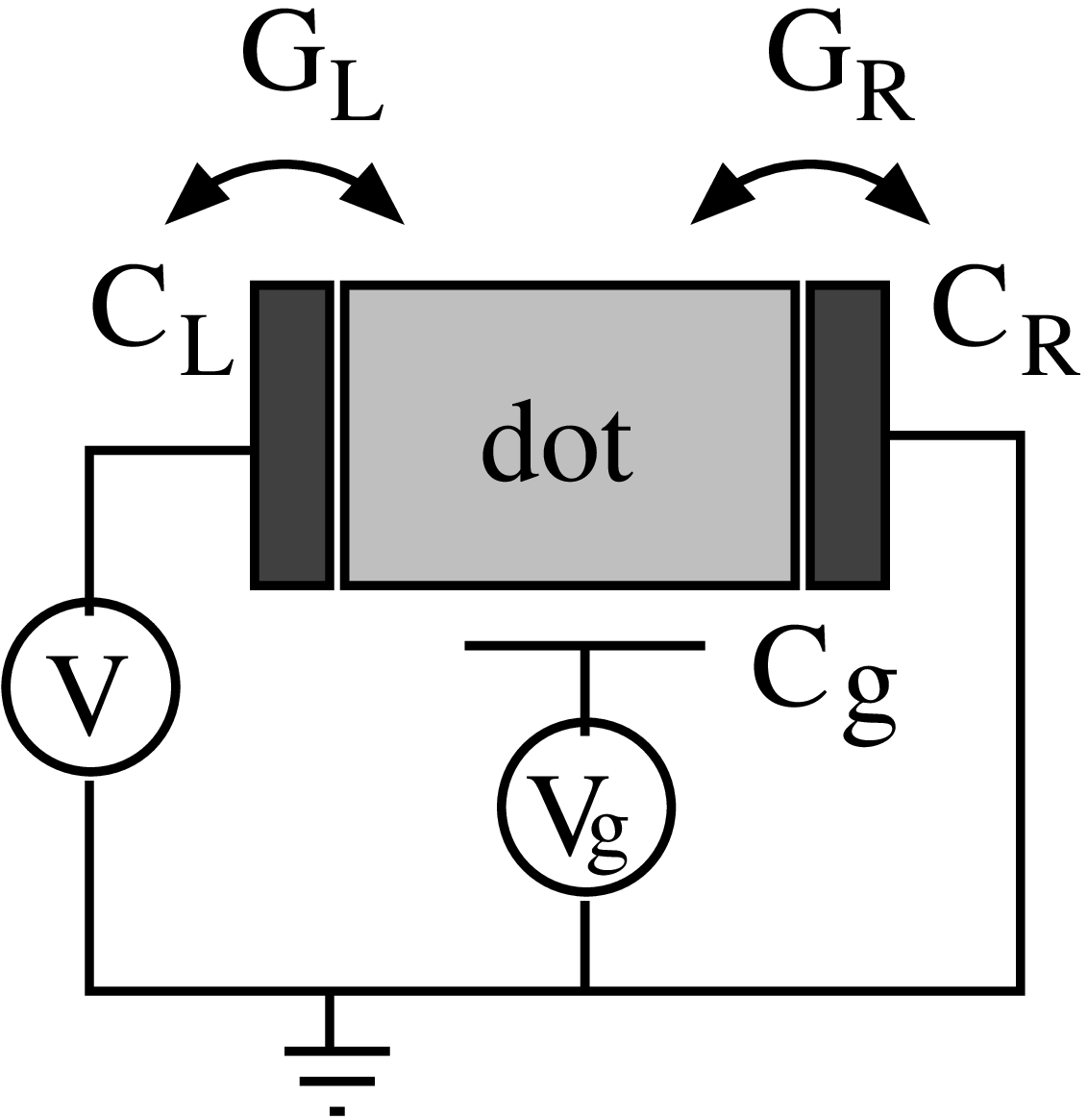}
\\
Fig.a
\\
\includegraphics[width=6cm]{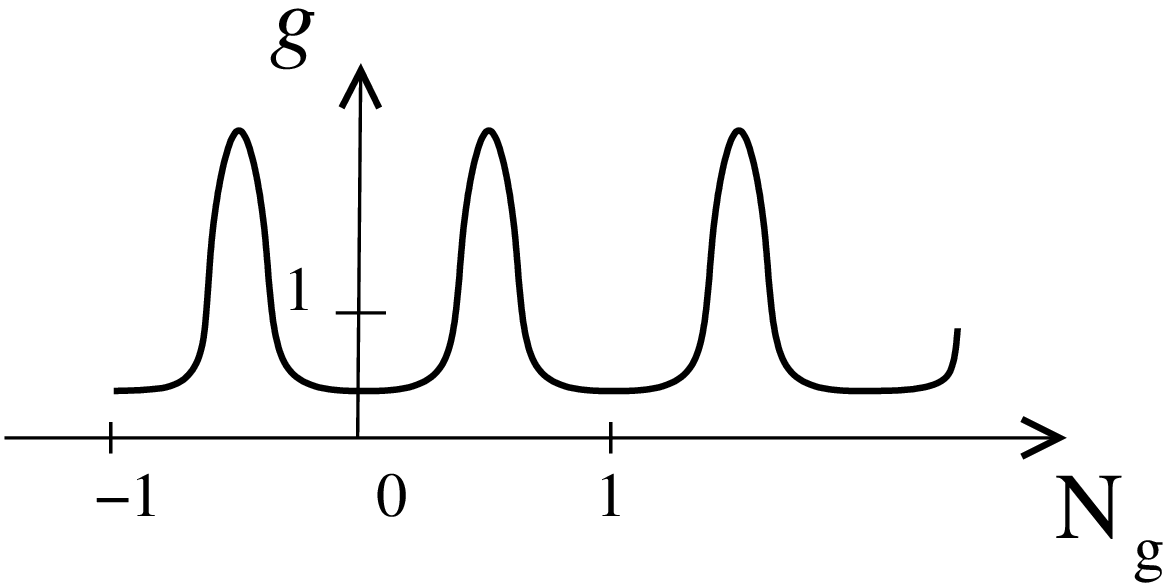}
\\
Fig.b
\end{center}
\caption{\label{fig:SET}
(a) Sketch of a single electron transistor. (b) Conductance of the SET as a function of gate 
voltage for $E_C \ll T \ll \Delta$. At the degeneracy points  Coulomb blockade is lifted 
and transport is allowed through  the single electron transistor. }
\efig

Let us first study the conductance of the SET in the regime where the difference $\Delta E$ 
between the energy of the two charging states considered is much larger than the 
temperature. It turns out that the range of validity of Eq.~(\ref{activated}) describing an
activated behavior is rather small for typical parameters, 
and  the conductance is dominated  by second order 
virtual processes as soon as we lower the temperature much below $\Delta E$. 
From the point of view of these  second order processes two regimes must be distinguished: 
In the regime  $\Delta \ll T \ll E_C$ the leading term to the conductance comes from 
elastic and inelastic  {\em  co-tunneling} processes
shown in Fig.~\ref{fig:cotun}:\cite{cotunneling} 
In the inelastic co-tunneling  process a conduction electron jumps into the 
dot from one lead and another electron jumps out of the dot to the other lead 
in a second order virtual process, leaving behind 
(or absorbing) an electron-hole excitation on the dot, 
while in an elastic co-tunneling it is the same electron that jumps out.
 Inelastic co-tunneling gives a conductance  $G_{\rm inel} \sim G_L G_R (T/E_C)^2/ (e^2/h) $, 
and is thus 
clearly suppressed as the temperature decreases,\cite{cotunneling} 
while elastic co-tunneling 
results in a small temperature independent residual conductance,
 $G_{\rm el} \sim G_L G_R (\Delta/E_C)/ (e^2/h) $
even at $T=0$ temperature.\cite{cotunneling}

\bfig
\begin{center}
\includegraphics[width=5cm]{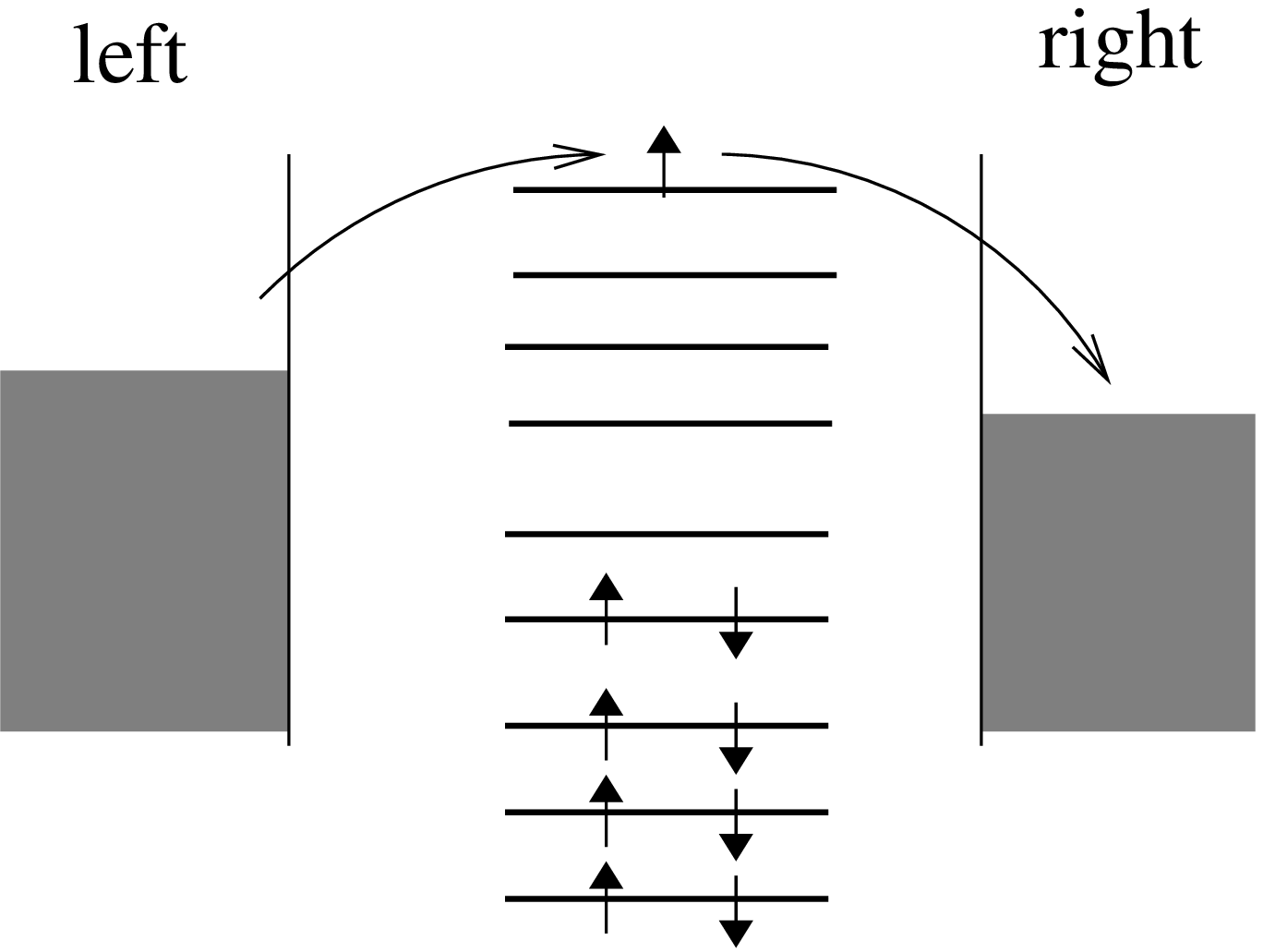}
\\
Fig.a
\\
\includegraphics[width=5cm]{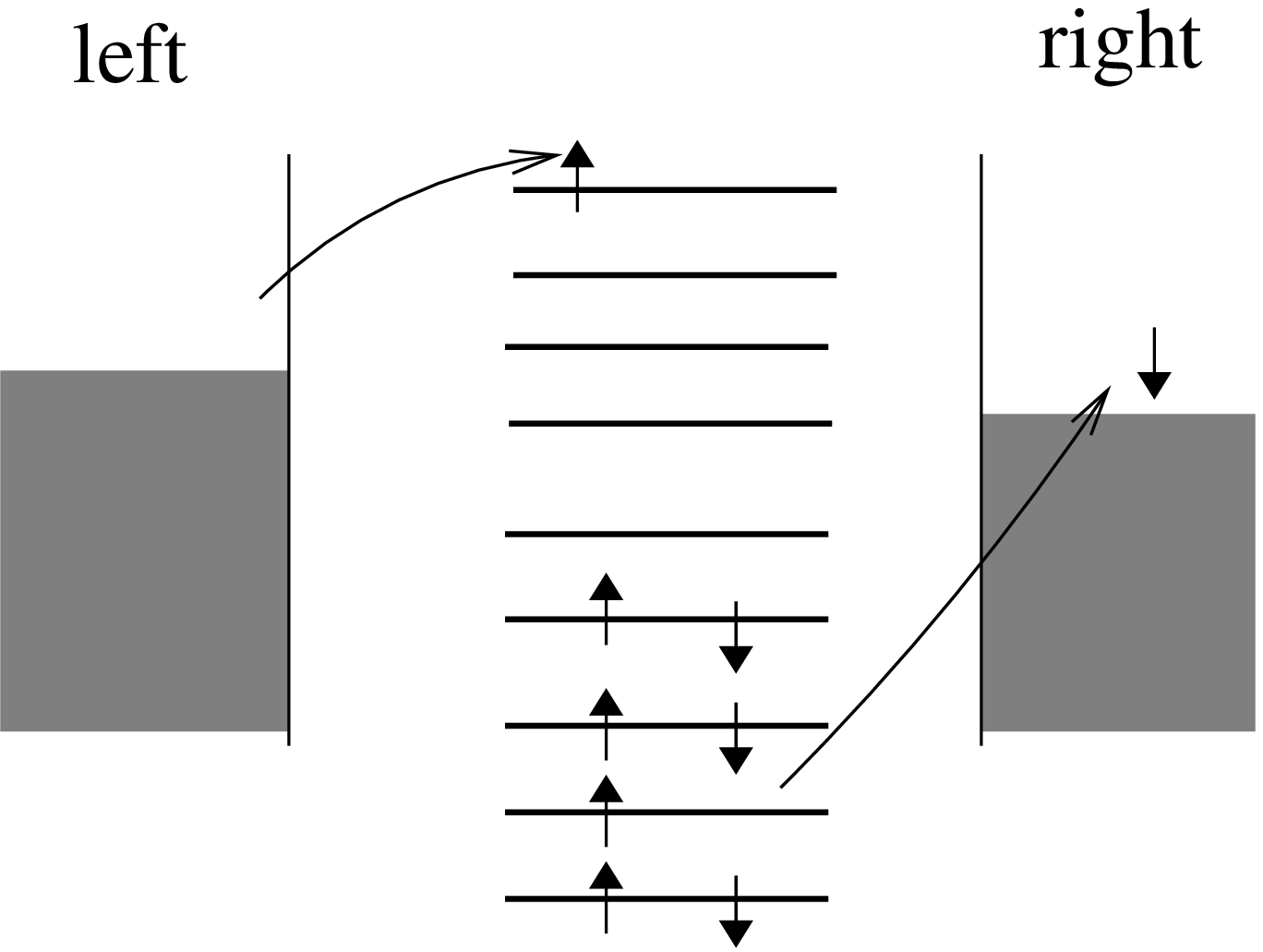}
\\
Fig.b
\end{center}
\caption{\label{fig:cotun} 
Elastic (a) and inelastic (b) co-tunneling processes. Inelastic co-tunneling processes
give a conductance $\sim T^2$ while elastic co-tunneling gives a finite conductance
as $T\to 0$.
}
\efig

\begin{figure}[thb]
\centering
\includegraphics[width=6cm]{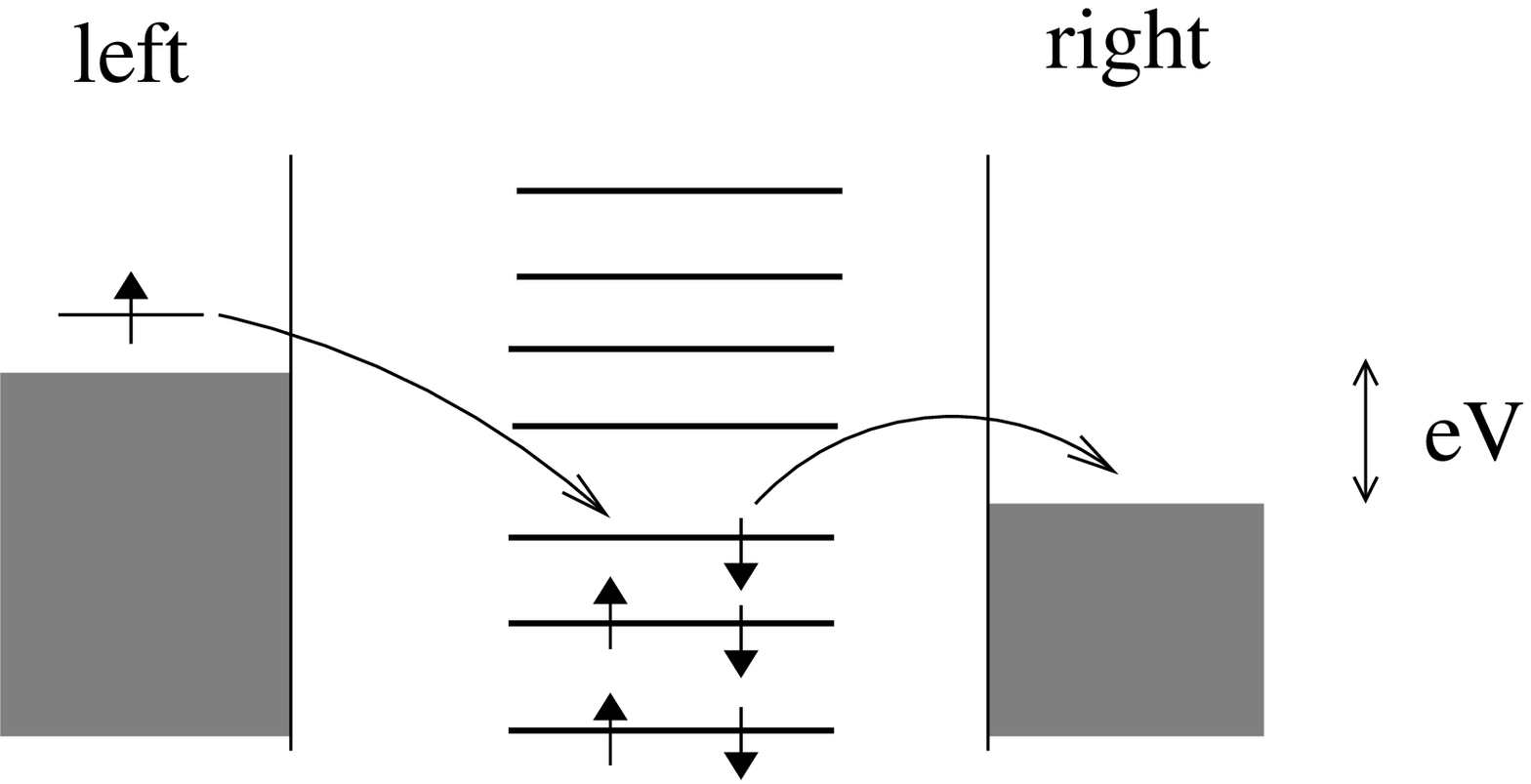}
\caption{\label{fig:exchange} Exchange process leading to the enhancement of the conductance as
$T\to0$.}
\efig
\section{Kondo effect}
\label{sec:Kondo}
For $T \ll \Delta   \ll E_C$ inelastic co-tunneling processes  are not allowed,
and the properties of the SET depend essentially on the {\em number of electrons} 
on the dot. The ground state of the isolated dot must be 
spin degenerate if there is an {\em odd number of electrons} on the 
dot, while it is usually non-degenerate, if the number of electrons on the dot is even.  
In the latter case nothing special happens: quantum fluctuations due to 
coupling to the leads produce just a residual conductance as $T\to 0$. 
If, however, the number of the electrons on the dot is {\em odd},
then the ground state has a {\em spin degeneracy}, which can give rise to 
the Kondo effect discussed below. In this case, exchange processes
shown in Fig.~\ref{fig:exchange} give a contribution to the conductance. 
As we lower the temperature, the effective amplitude of these processes increases
due to the Kondo effect, and ultimately gives a conductance  that can be as large 
as the  quantum conductance $G_Q = 2e^2/h$ at $T=0$ temperature.\cite{Leo} 
This strong enhancement is due to strong quantum fluctuations of the spin of the dot, 
and the formation  of a strongly correlated Kondo state. The typical temperature dependence 
of the conductance  for $N_g$ odd is shown in  Fig.~\ref{fig:g(T)}.

\begin{figure}[bht]
\centering
\includegraphics[width=6cm]{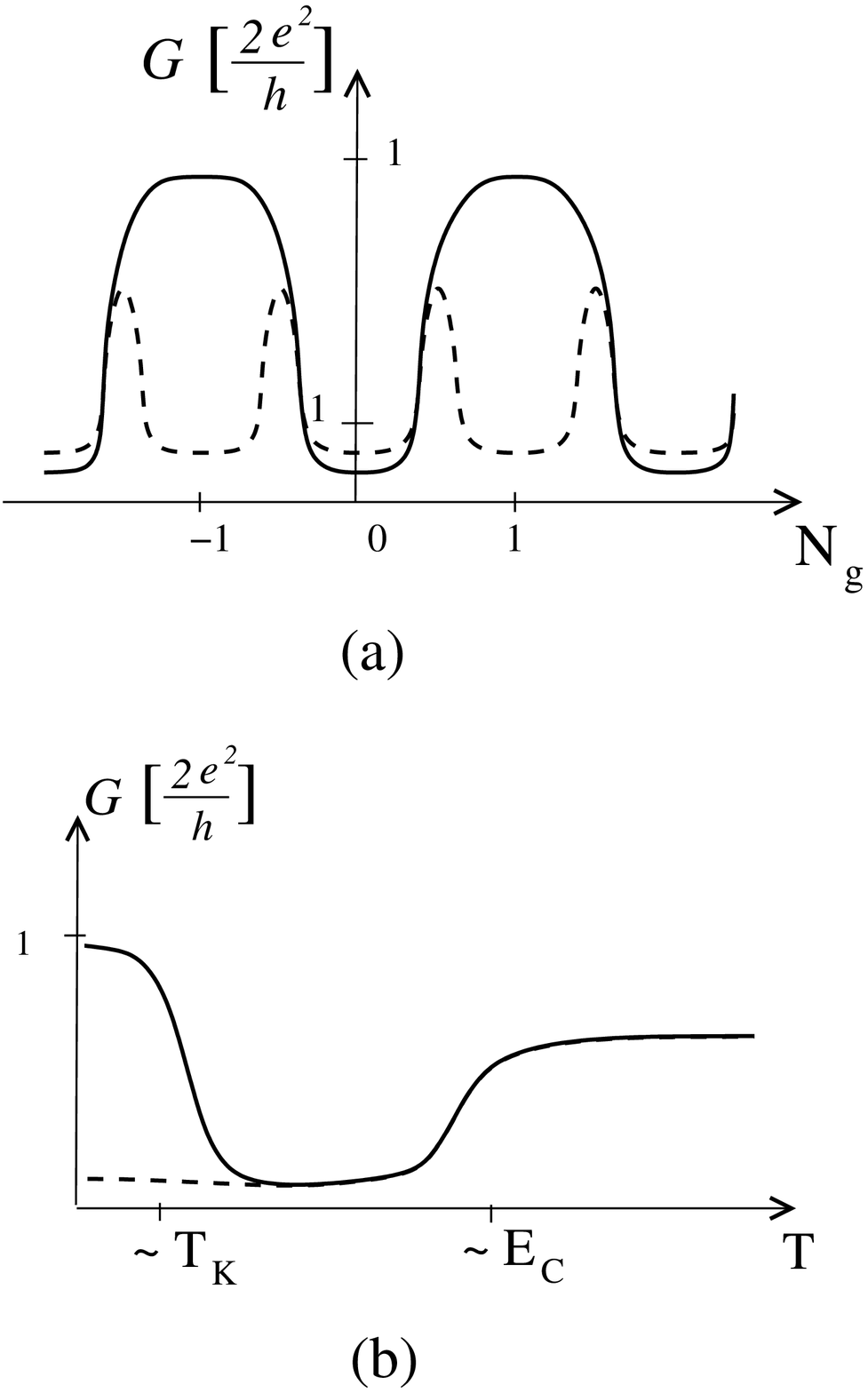}
\caption{\label{fig:g(T)} (a) Linear conductance of the SET for $T_K\ll T\ll E_C$ 
(dashed line) and $T\to 0$ (solid line). (b)  Temperature dependence of the conductance 
of a symmetrical SET for an odd number of electrons on the dot (continuous line) and for an even number 
of electrons with non-degenerate ground state (dashed line).}
\end{figure}

To understand why the conductance of the dot becomes large, 
let us keep only the last, singly occupied  level that gives rise to the Kondo effect, 
 $d_{j\sigma}\to d_\sigma$, $\epsilon_j \to \epsilon_d$, 
and write the  Hamiltonian of the dot as
\bea 
H & =&  \sum_{\sigma} \epsilon_d \;d^\dagger_{ \sigma } d_{\sigma } 
+ H_C 
\nonumber
\\
&+&    
\sum_{\sigma} \sum_{\mu=L,R} \int {d\epsilon}\; \epsilon \;
\psi_{\mu,\sigma}^\dagger(\epsilon) \psi_{\mu,\sigma}(\epsilon)
\\
\nonumber
&+&    
  \int d\epsilon \; \Bigl\{t^L \varrho_L^{1/2} \;  (d^\dagger_{\sigma } \psi_{L, \sigma}(\epsilon) 
+ {\rm h.c.}) + ``L\leftrightarrow R''\Bigr\}\;, 
\eea
where we introduced the fields, 
$
\psi_{L,\sigma}(\epsilon) \equiv \varrho_{L/R}^{1/2} \; \psi_{L/R, \epsilon \sigma}
$.
For the sake of simplicity, let us assume that  
$t^R \varrho_L^{1/2}  = t^L \varrho_R^{1/2}$.
If we then make a unitary transformation and introduce the even and odd 
field operators, 
\be
\psi_\pm \equiv {\psi_R \pm \psi_L\over \sqrt{2}}\;,
\label{unitary}
\ee
then obviously  the odd combination  $\psi_-$ fully decouples from the dot,
and the tunneling part of the Hamiltonian can be written as 
\be 
\hat V =  \tilde t \int d\epsilon \; (d^\dagger_{\sigma } \psi_{+, \sigma}(\epsilon) 
+ {\rm h.c.}) \;,
\ee
where $\tilde t = \sqrt{2} \; t^R \varrho_R^{1/2}$.  
One can now perform second order degenerate perturbation theory 
in $\hat V$ in the subspace $\sum_\sigma d^\dagger_\sigma d_\sigma = 1$ 
to obtain the following effective exchange Hamiltonian:
\bea 
H_{\rm eff} & = & 
\sum_{\sigma}  \sum_{\epsilon}\; \epsilon \;
\psi_{+, \sigma }^\dagger(\epsilon) \psi_{+,\sigma}(\epsilon) 
\label{eq:exchange}
\\
\nonumber
&+& {J\over 2} {\vec S} \; \sum_{\sigma,\sigma'}\int\int d\epsilon\;d\epsilon' 
\psi_{+\sigma}^\dagger(\epsilon) \vec \sigma_{\sigma\sigma'} 
\psi_{+\sigma'}(\epsilon') \;,
\eea
where $\vec S = {1\over 2} (d^\dagger \vec \sigma d)$ is the spin  
of the dot, and $J\sim \tilde t^2/E_C$ is a dimensionless antiferromagnetic 
exchange coupling.   Thus  electrons in the  even channel
 $\psi_{+}$  couple antiferromagnetically  to the spin on 
the partially occupied 
$d$-level, and try to screen it to get rid of the residual entropy associated 
with it.

Written 
in the original basis,  Eq.~(\ref{eq:exchange}) contains terms 
$\sim J  {\vec S} 
\psi_{R}^\dagger(\epsilon) \vec \sigma \psi_{L}(\epsilon')$,
which allow for charge transfer from one side of the dot to the other side,
and in leading order these terms give a conductance $\sim J^2$. However, higher order terms 
in $J$ turn out to give logarithmically divergent  contributions,
\be
G\sim {e^2\over h} \Bigl(J^2 + 2\;J^3\;{\rm ln}(\Delta/T) + \dots \Bigr)\;.
\label{eq:G_log}
\ee
As a result, the conductance of the device increases as we decrease 
the temperature and our perturbative approach  breaks down at the
 so-called Kondo temperature, 
\be 
T_K \approx \Delta e^{-1/J}\;.
\ee
One can try to get rid of the logarithmic singularity
in Eq.~(\ref{eq:G_log}) by summing up the most singular contributions in each 
order in $J$. This can be most easily done by performing a renormalization 
group calculation and replacing  $J$ by its renormalized value in the 
perturbative expression,  $G\sim {e^2\over h} J^2$.\cite{RG}
However, this procedure does not cure the problem 
and gives a conductance that still diverges at $T=T_K$,
\be
G\sim  {e^2\over h} {1\over {\rm ln}^2(T/T_K)}\;.
\ee
The meaning of the energy scale $T_K$ is that below this temperature scale 
the effective exchange coupling becomes large and  a conduction 
electron spin is tied antiferromagnetically to the spin of the dot to form a singlet
(see Fig.~\ref{Kondo_effect}).

\bfig[b]
\includegraphics[width=5cm]{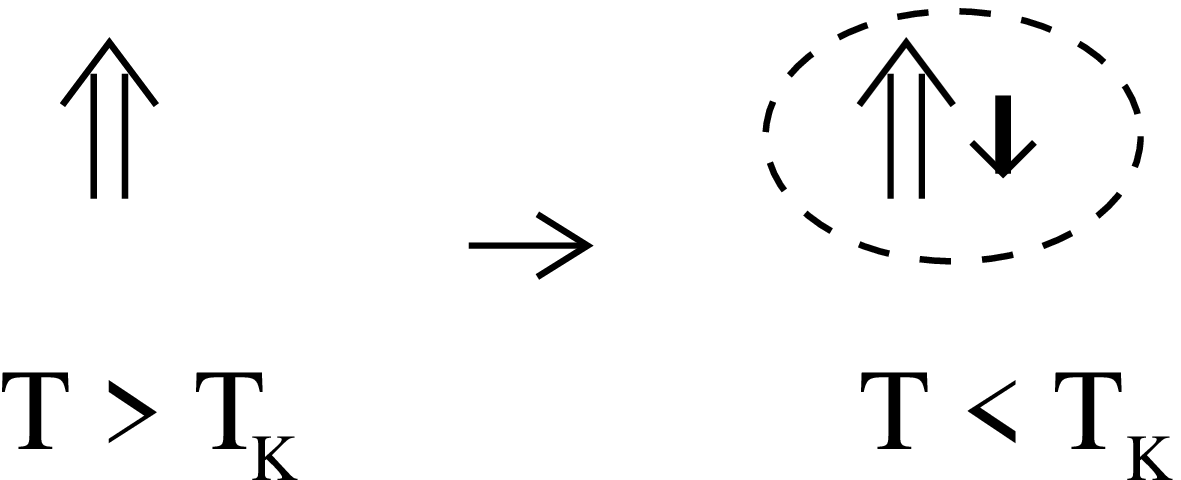}
\caption{
\label{Kondo_effect} 
Conduction electrons in the even channel screen the impurity spin. 
}
\efig

It is not difficult to show that then for $t^L\varrho_L^{1/2} = t^R\varrho_R^{1/2} $
the SET must have perfect conductance  at $T=0$ temperature.  To show this, let us apply 
the Friedel sum  rule,\cite{Hewson,Friedel_Kondo} 
that relates the number $N_{\rm bound}$  of electrons 
bound to the impurity and the phase shifts $\delta$ of the electrons $\psi_+$, as
\be
N_{\rm bound} = 2 {\delta\over \pi}\;,
\ee
where the factor $2$ is due to the spin.  
This relation 
implies that in the even channel conduction electrons acquire a phase shift $\delta = \pi/2$, 
and correspondingly, $\psi_+ \to e^{2i\delta} \psi_+ = - \psi_+$ in course of a scattering 
process at $T=0$ temperature. 
Going back to the original left-right basis, this implies that left and right electrons scatter 
as
\be 
\psi_L \to -\psi_R\;,\phantom{nnn} \psi_R \to -\psi_L\;.
\ee
In other words, an electron coming from the left is transmitted 
to the right {\em without any backscattering}, and thus the quantum dot has a perfect conductance,
$2e^2/h$. 

We remark here that only a symmetrical device can have a perfect transmission, and only if 
the the number of electrons on the $d$-level is approximately one, 
$\langle n_d\rangle \approx 1$. All the considerations above can be easily generalized to the 
case $t^L\varrho_L^{1/2} \ne t^R\varrho_R^{1/2}$, and one obtains for the zero temperature 
conductance in the Kondo limit, 
\be
G(T\to0) = {2e^2\over h} {4 \varrho_L t_L^2 \;\varrho_R t_R^2 \over 
( \varrho_L t_L^2 + \varrho_R t_R^2 )^2}\;,
\ee 
which is clearly less than $2e^2/h$ for non-symmetrical dots.

\begin{figure}[tb]
\centering
\includegraphics[width=6cm]{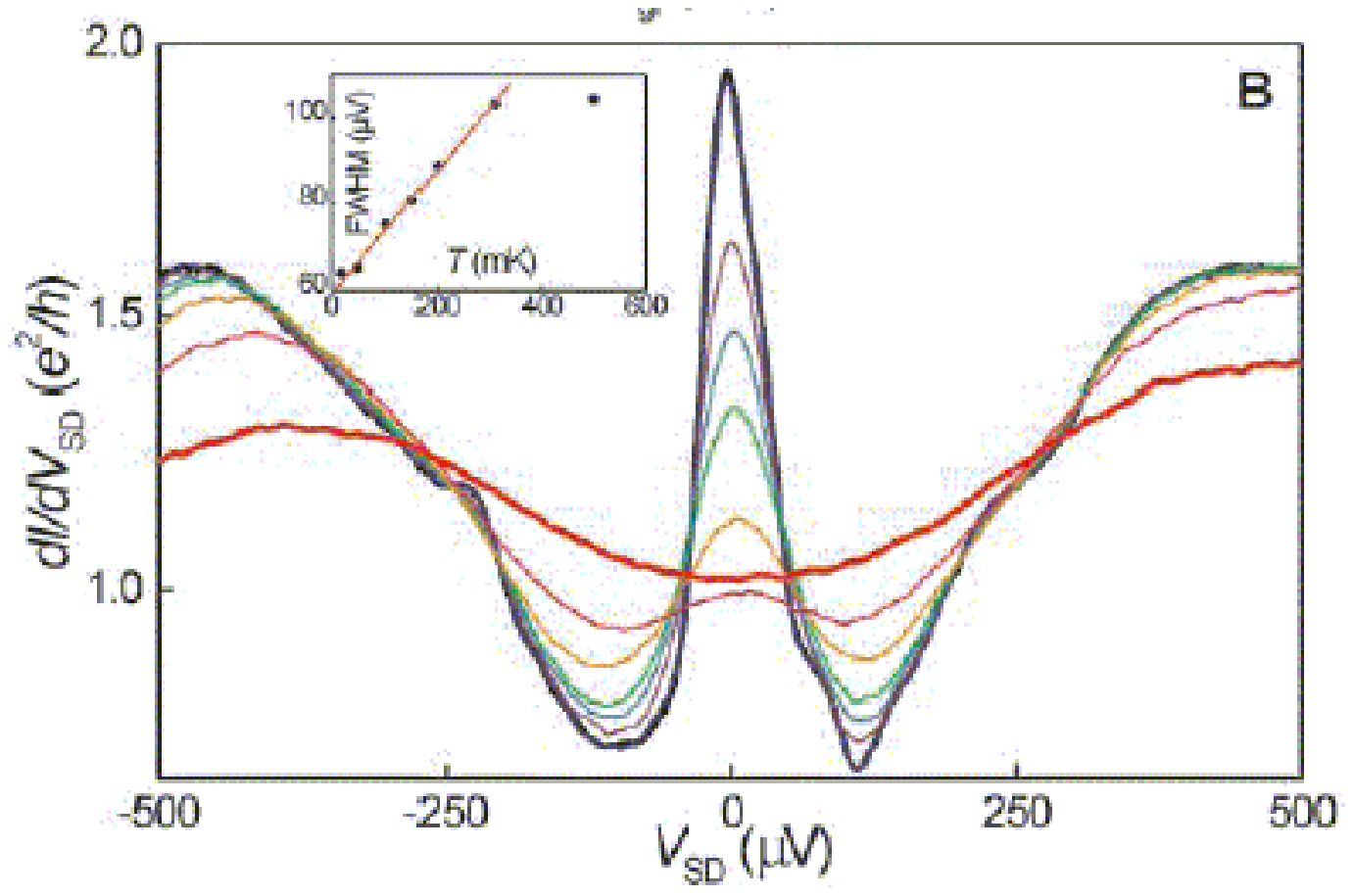}
\caption{
\label{fig:leo}
The Kondo resonance appears as a peak in the differential conductance of the 
SET (figure taken from Ref.~\onlinecite{Leo}). At $T=0$ temperature the conductance approaches the quantum limit of the 
conductance, $2e^2/h$. 
}
\end{figure}

The  phase shift $\delta = \pi/2$ also implies that there must be a {\em resonance} 
at the Fermi energy. 
In fact, this resonance is called the Kondo resonance, and can be directly seen in 
the differential conductance (related to the density of states as usual) 
of the single electron transistor shown in Fig.~\ref{fig:leo}. 
This is a many-body resonance that develops at the Fermi energy (zero bias)
as the temperature is cooled down below the Kondo  temperature $T_K$.

The basic transport properties of the SET have been  summarized in Fig.~\ref{fig:g(T)}.
Although we could get  a fairly good analytical understanding of behavior of a SET, 
based on the simple considerations outlined above, 
it is rather difficult to obtain a {\em quantitative} description. In fact, 
 to obtain a quantitative description extensive computations such as numerical renormalization 
group calculations are needed \cite{Theo}. A valid and complete description of 
the out of equilibrium physics of a SET is still missing.\cite{Meir,Achim,Piers,Schiller}

\section{Orbital degeneracy and correlations in quantum dots}
\label{sec:orbital}

In the previous section we sketched the generic behavior of a quantum dot, and assumed 
that the separation of the last, partially occupied level is by an energy distance 
 $\sim \Delta\gg T$ separated 
from all other single particle levels of the dot. This is, however, not always true. 
In the case of a symmetrical arrangement
like the one shown in Fig.~\ref{fig:triangle}, {\em e.g.},   
some of the states of the dot are orbitally degenerate 
by {\em symmetry},\cite{Sasaki,triangle} 
while in some other cases almost degenerate orbitals may  show up 
just by accident.\cite{vanderWiel,KoganST} 
This orbital degeneracy can play a very important role when we fill up these 
degenerate (or almost degenerate) levels, and leads to such phenomena as 
the SU(4) Kondo state \cite{triangle,su4}  or eventually the singlet-triplet 
transition.\cite{vanderWiel,Sasaki,KoganST}

\begin{figure}[htb] 
\centering
\includegraphics[width=7cm]{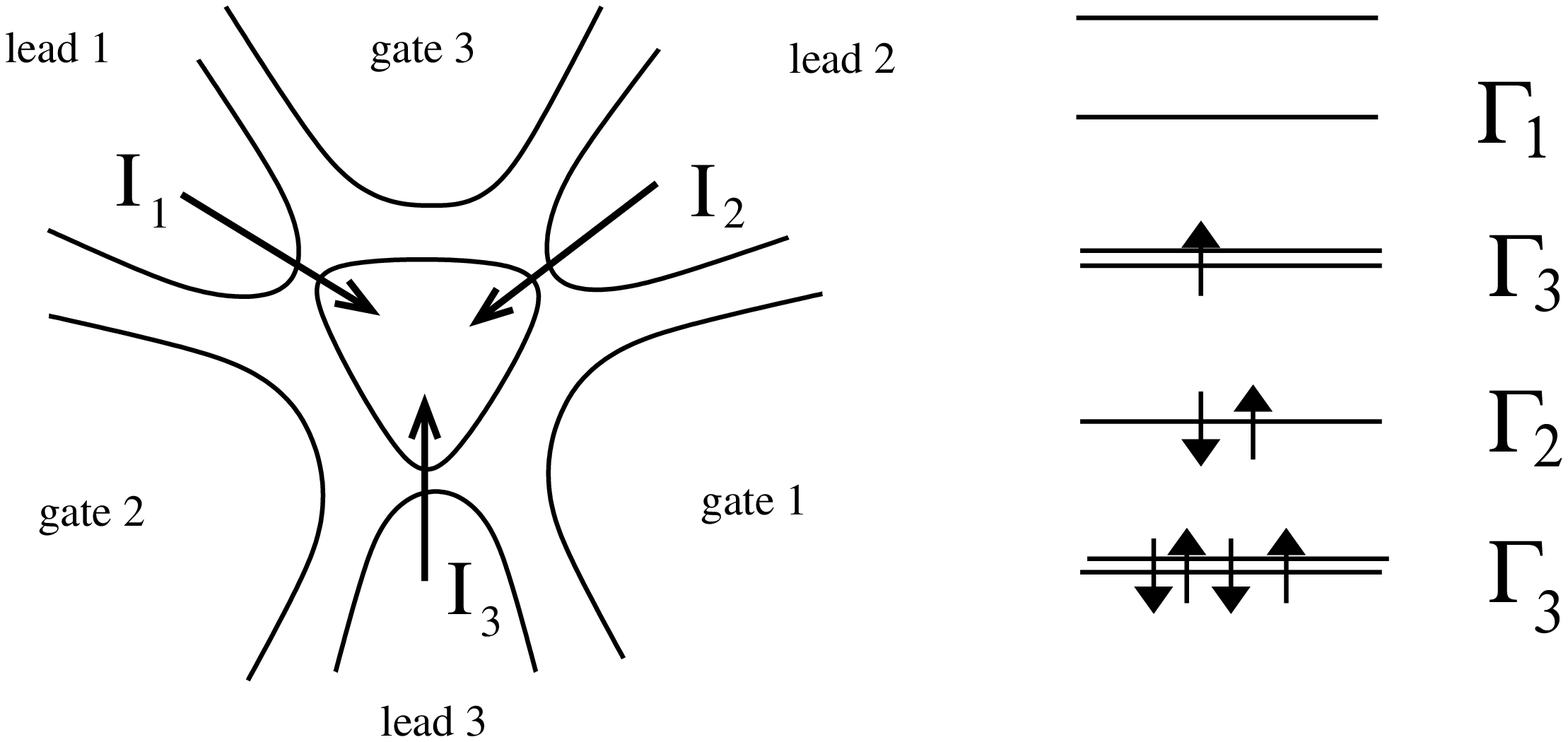} 
\caption{Arrangement with triangular symmetry
and the structure of the four-fold  degenerate ground state of the 
isolated triangular dot.} 
\label{fig:triangle} 
\end{figure}

\vskip0.2cm
{\bf
\parindent=0pt 
 SU(4) physics and triangular dots} 
\vskip0.2cm

Let us first discuss what happens if we have just a single electron 
on an orbitally degenerate level. 
The presence of the orbital degeneracy leads to an unusual state of 
(approximate) SU(4) symmetry in this case, where spin and charge degrees of freedom are 
entangled. This SU(4) state has been proposed first theoretically to appear in double dot systems 
and quantum dots with triangular symmetry in Refs.~\onlinecite{su4}
and \onlinecite{triangle}.
 However, while there is no unambiguous experimental evidence of an SU(4) 
Kondo effect in double dot devices,\cite{Weis} 
the SU(4) state has been recently observed in vertical dots 
of cylindrical symmetry\cite{Sasaki2} as well as in carbon  nanotube 
single electron transistors.\cite{Herrero} In both chases the degeneracy index 
is due to a chiral symmetry just as in Ref.~\onlinecite{triangle}. 
Other realizations of states with SU(4) symmetry have been also proposed later in 
more complicated 
systems,\cite{Pascalsu4,LaciPascalsu4} and also in the context of 
heavy fermions.\cite{Michelesu4}
 
For the sake of concreteness, we shall focus here to the case of the triangular 
dot shown in Fig.~\ref{fig:triangle}. However,  our discussions carry over 
with trivial modifications to the previously mentioned experimental systems 
in Refs.~\onlinecite{Sasaki2} and \onlinecite{Herrero}. 
Let us first assume that we have two orbitally degenerate levels $|\pm\rangle$
that can be labeled by some chirality index $\tau = \pm$, and let us focus on the 
charging of this multiplet only. At  the Hartree-Fock level,  these levels of the   isolated dot  
can be  described by:
\begin{eqnarray}
H_{\rm dot}  =  \sum_{\tau,\tau',\sigma} 
d^\dagger_{\tau \sigma}
(E_{\tau \tau'} + \Delta E \;\delta_{\tau \tau'}) 
d_{\tau' \sigma} 
-  {J_H\over 2} {\vec S}^{\;2}  \nonumber  \\
 +  {E_C \over 2} (n_+ + n_-)^2 
+ {\tilde E_C\over 2} (n_+ - n_-)^2\;, 
\label{eq:H_dot} 
 \end{eqnarray}
where $d^\dagger_{\tau \sigma}$ creates an electron on the dot
within the degenerate multiplet 
with spin $\sigma$ and orbital label $\tau$. The energy shift 
$\Delta E$ above is proportional to the (symmetrically-applied) 
gate voltage and controls the charge on the dot,
while $E_{\tau \tau'}$ accounts for the splitting generated 
by  deviations from perfect triangular symmetry ($\sum_\tau E_{\tau \tau }=0$).
We denote the total number of electrons in state $\tau = \pm$ by 
$n_\tau \equiv \sum_\sigma d^\dagger_{\tau \sigma}d_{\tau \sigma}$,
and ${\vec S} = {1\over 2} \sum_{\tau,\sigma,\sigma'} 
d^\dagger_{\tau \sigma}{\vec \sigma}_{\sigma \sigma'} d_{\tau \sigma'}$
is the total spin of the dot. The terms proportional to $E_C$ and 
${\tilde E_C}$ are generated by the Hartree interaction, while
that proportional to $J_H$ in  Eq.~(\ref{eq:H_dot}) is the
Hund's rule coupling, generated by exchange. This term has no importance
if there is only a single electron on the dot.

\begin{figure}[htb] 
\centering
\includegraphics[width=8cm]{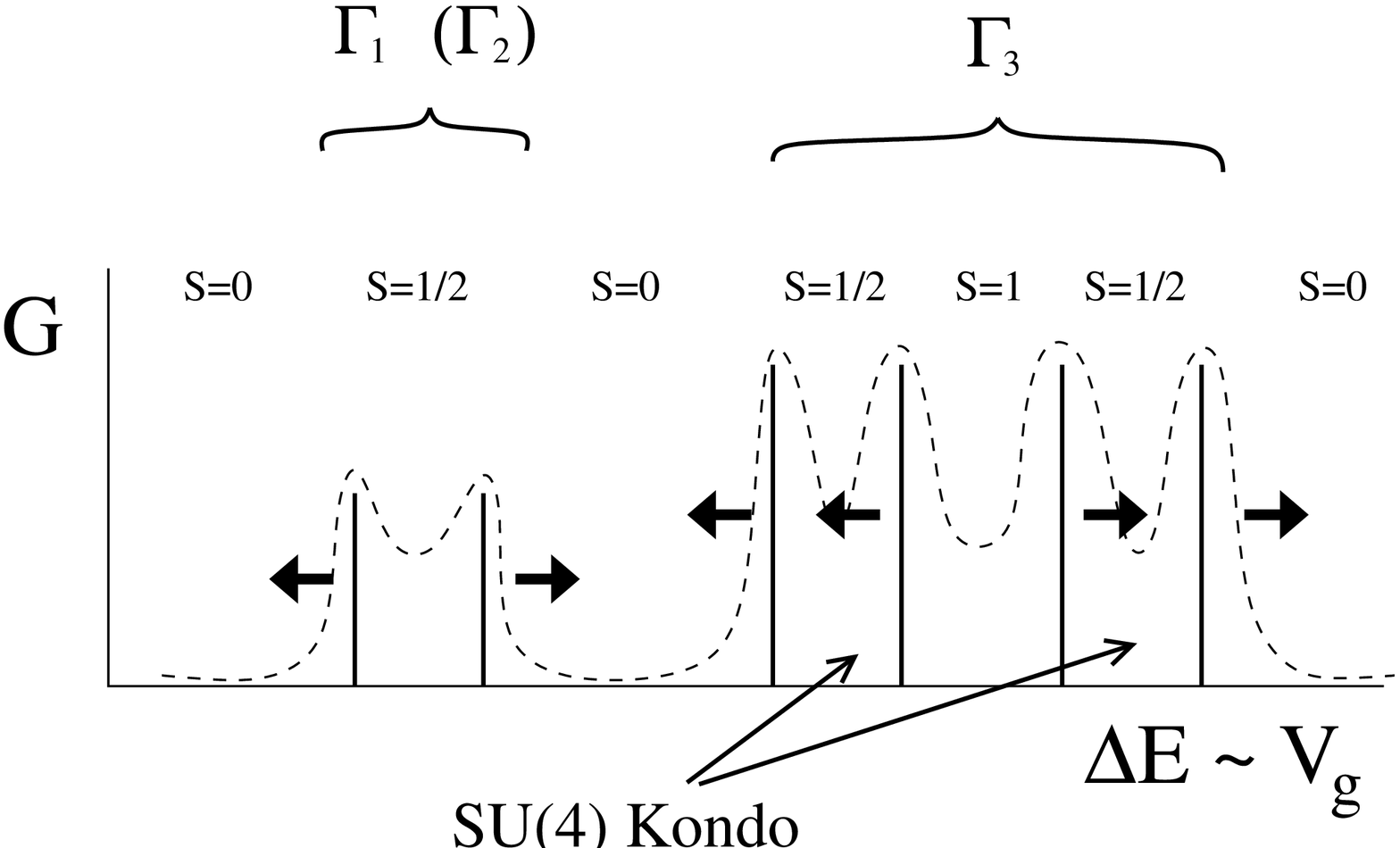}
\caption{Structure of the Coulomb blockade peaks of the triangular dot.
The multiplets are labeled by the corresponding irreducible representations. The arrows 
indicate the direction the peaks move when applying a Zeeman field.}
\label{fig:Coulomb}
\end{figure} 
Let us first consider the linear conductance.  The currents $I_j$ between leads $j$ and the dot 
are related to the voltages $V_j$ applied on them by the conductance tensor,   
\be
I_j =  \sum_{j'} G_{jj'} V_{j'}\;,
\ee
which further simplifies to   $G_{j j'} =  {3\over 2}  G \delta_{j j'} - G/2$
for a symmetrical system.    A schematic 
plot of the conductance  $G$ 
as a function of $\Delta E$  is shown in Fig.~\ref{fig:Coulomb}.  
The arrangement of  the four Coulomb blockade peaks associated to the 
fourfold degenerate $\Gamma_3$ state is symmetrical, and  the height of the four 
peaks turns out to be numerically almost identical at high temperatures.\cite{triangle} 

The most interesting regime in Fig.~\ref{fig:Coulomb}
appears between the first two peaks. Here there is 
one electron on the dot, and correspondingly the ground state of the 
isolated dot is fourfold degenerate. Let us now tunnel couple the dots 
to conduction electrons in the three leads, $\psi_j$ ($j=1,2,3$). To simplify the 
Hamiltonian  we first introduce new fields that transform with the same symmetry as 
the states $|\tau\rangle$, 
\be 
\psi_j \to \psi_\pm \equiv {1\over \sqrt{3}} \sum_j e^{\pm i\;2\pi j
/3} \psi_j\;.
\ee
With this notation the tunneling Hamiltonian of a perfectly symmetrical dot 
becomes
\be
\hat V =   t \sum_{\sigma,\tau} \int d\epsilon\; \Bigl( d_{\tau \sigma}^\dagger
  \psi_{\tau \sigma}(\epsilon)  + h.c.\Bigr)\;.
\label{eq:V_triangle}
\ee

To describe the fourfold degenerate ground state  of the dot, 
we can introduce the spin 
and and orbital spin operators $\vec S$ and $\vec T$,
with $S_z = \pm1/2$ and $T_z=\pm 1/2$ corresponding to the states
$\sigma=\pm$ and $\tau = \pm$.  Likewise, we can introduce spin and 
orbital spin operators  $\vec \sigma$ and $\vec \tau$ also for the conduction electrons,
and then proceed as in the previous section to generate an effective Hamiltonian 
by performing second order perturbation theory in the tunneling 
$\hat V$, Eq.~(\ref{eq:V_triangle}).  
The resulting interaction Hamiltonian  is rather complex 
and contains all kinds of orbital and spin couplings
of the type $\sim  T^+ \tau^-  {\vec S} {\vec \sigma}$
or $\sim T^z \tau^z $.\cite{su4,triangle} These terms are 
generated by processes like the one shown in Fig.~\ref{fig:spin-orbit_coupling},
and clearly couple spin and orbital fluctuations to each-other. 

\bfig[tb]
\includegraphics[width=5cm]{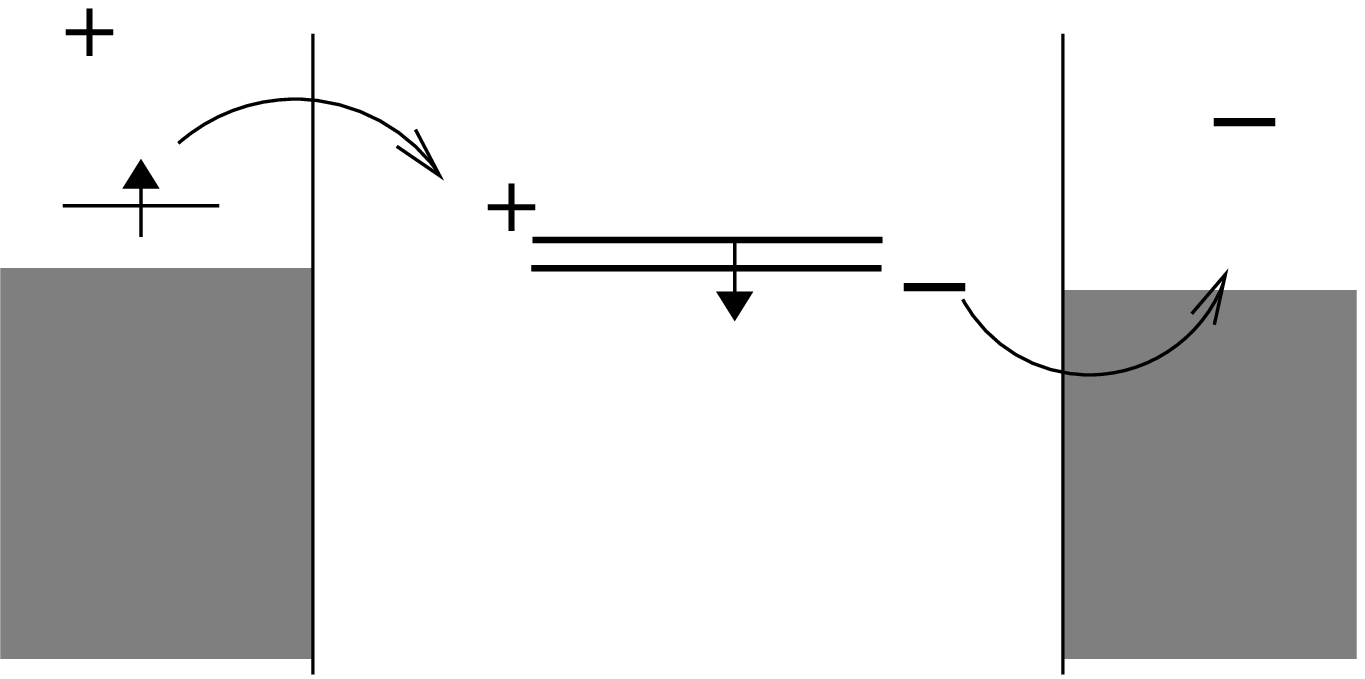}
\caption{
\label{fig:spin-orbit_coupling}
Example of a virtual process generating a coupling between the spin and
the orbital spin. 
} 
\efig 

Fortunately, a renormalization group 
analysis reveals that at low temperatures the various couplings become equal, and 
the Hamiltonian can be simply replaced by the following 
remarkably  simple  SU(4) symmetrical  effective Hamiltonian 
(Coqblin-Schrieffer model),
\begin{equation}
H_{\rm eff}(T\to0) = \tilde J \sum_{\alpha,\beta = 1,..,4} \psi_\alpha^\dagger 
\psi_\beta \; |\beta\rangle \langle \alpha|\;,
\label{eq:H_eff}
\end{equation}
where the index $\alpha$ labels the four combinations of 
possible spin and pseudospin  indices, and the $|\alpha\rangle$'s 
denote the four states of the dot.  The dynamical generation of this 
SU(4) symmetry  can also be verified by 
solving the original complicated Hamiltonian by the powerful machinery of 
numerical renormalization  group.\cite{Wilson,su4}
We remark here that the structure  of the fixed point 
Hamiltonian,  Eq.~(\ref{eq:H_eff}) is rather robust.  Even if the 
system does not have a perfect triangular (or chiral) symmetry, 
the exchange part of the effective Hamiltonian at low temperatures 
will take the form  Eq.~(\ref{eq:H_eff}),  and the effect of imperfect symmetry 
only generates some splitting $E_{\tau\tau'}$ for the orbitally degenerate levels 
and some small potential scattering.\cite{Mstate}  These terms, of course, break the 
SU(4) symmetry of  Eq.~(\ref{eq:H_eff}), but represent only marginal perturbations, 
and do not influence the physical properties of the system 
in an esential way if they are small.
In a similar way, the SU(4) symmetric fixed point discussed here 
may be relevant even for systems with (approximate) accidental degeneracy
even if they do not have a perfect SU(4) symmetry.

Similar spin and orbital entangled states apparently also show up in molecular clusters, but 
there they may lead to the appearance of unusual non-Fermi liquid 
states.\cite{Crommie,Cr3,Ingersent}

\begin{figure}[bt]
\includegraphics[width=5cm]{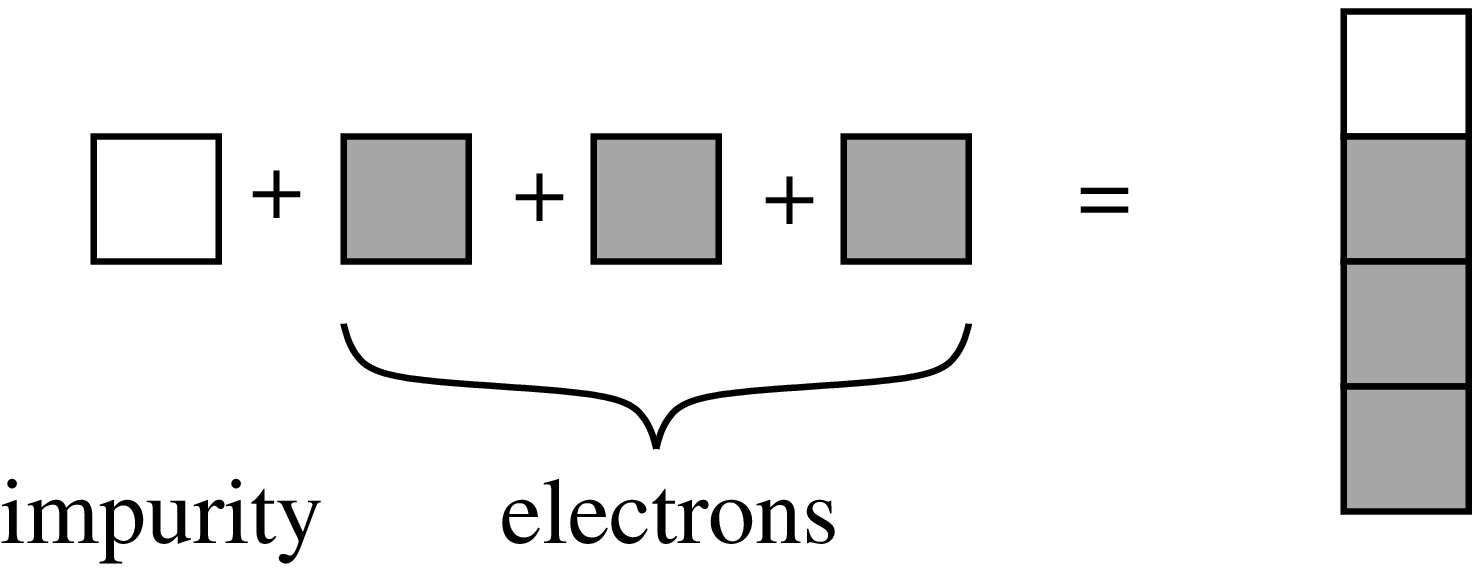}
\caption{
\label{fig:SU4_screen}
To screen the SU(4) spin of a triangular dot one needs three conduction 
electrons. The singlet formed corresponds to the Young tableau on the right, while
the defining four-dimensional SU(4) representations are denoted by squares.
} 
\end{figure}

The Hamiltonian Eq.~(\ref{eq:H_eff}) is one of the exactly solvable  
models,\cite{WiegmannTsvelick} and has been  studied thoroughly 
before.\cite{CoqblinSchrieffer,NozieresBlandin} 
Just as in the Kondo problem studied in the previous section, the SU(4) spin of the 
dot is screened below the 'SU(4)' Kondo temperature, $T_K^{(0)}$.
However, to screen an SU(4) spin, 
one needs {\em three} conduction electrons, as shown in Fig.~\ref{fig:SU4_screen}.
As a result, the Friedel sum rule in the present case is modified to 
\be 
3 = \sum_{\sigma,\tau} {\delta_{\sigma\tau}\over \pi} = 4 {\delta\over \pi}\;,
\ee
corresponding to a phase shift $\delta = 3\pi/4 $ for the electrons 
$\psi_{\pm,\sigma}$. 

The application of   a magnetic field on the 
dot, $H\to H - B\; S^z$ clearly suppresses spin fluctuations.\cite{footnote_inplane} 
However, it does not suppress orbital fluctuations, which still lead to a 
more conventional SU(2) Kondo effect, by replacing the spin in the original 
Kondo problem. The Kondo temperature $T_K^{\rm orb}$ of this orbital Kondo effect is, however, somewhat 
reduced compared to the SU(4) Kondo temperature $T_K^{(0)}$,\cite{triangle} 
\be
T_K^{\rm orb} \sim {(T_K^{(0)})^2\over \Delta}\;.
\ee
The phase shifts in this case are simply $\delta_\uparrow \equiv \delta_{\pm,\uparrow} \approx \pi/2$, 
while $\delta_\downarrow \equiv \delta_{\pm,\downarrow} \approx  0$, just as for the original Kondo problem. 
The splitting of the two levels $\tau = \pm$ has a similar effect and drives the 
dot to a simple spin SU(2) state.

The zero-temperature phase shifts can be related to the transport properties of 
the device: From the  $T=0$ phase shifts $\delta_{\tau \sigma}$  
one can construct the  conduction electrons' scattering matrix in the original 
basis $\psi_j$ and compute all transport coefficients using the Landauer-Buttiker 
formula.\cite{Buttiker,triangle,Pustilnik} The $T=0$ conductance $G$ turns out 
to be independent of the magnetic field, and both for the SU(4) and orbital Kondo states one finds 
the same value, 
$$
G(B)= {8 e^2\over 9h}\sum_\sigma \sin^2(\delta_\sigma(B)) = {8 e^2\over 9h}\;.
$$ 
The {\em polarization} of the current, 
\be 
P = {\sin^2(\delta_\uparrow) - \sin^2(\delta_\downarrow) \over 
\sin^2(\delta_\uparrow) + \sin^2(\delta_\downarrow)}\;,
\label{eq:polarization}
\ee
however, {\em does} depend on the magnetic field and takes the values 
$P=0$ and $P=1$ in the SU(4) and orbital Kondo states, respectively. 
The phase shifts in Eq.~(\ref{eq:polarization}) can be extracted with very high 
accuracy from the finite size spectrum computed via the numerical renormalization 
group procedure,\cite{Wilson,su4,HofstetterZarand} and the results 
(originally computed for the  double dot system in Ref.~\onlinecite{su4})
are shown in Fig.~\ref{fig:filter}. 
Clearly, this device can be used as a { spin filter}: Applying a Zeeman 
field one can induce  a large spin polarization at low temperatures 
while having a large $\sim e^2/h$ conductance through the device.
A slightly modified version of this spin filter has indeed been 
realized in Ref.~\onlinecite{Herrero}, where two orbital states 
originating from different multiplets have been used to generate 
the orbital Kondo effect.

\begin{figure}[htb] 
\centering
\includegraphics[width=6cm]{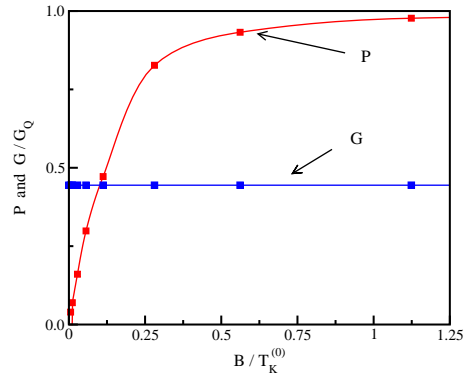}
\caption{\label{fig:filter} 
Spin polarization of the current through the triangular dot as a function 
of magnetic field. }
\end{figure}

To close the analysis of the SU(4) Kondo effect, let us shortly 
discuss how the SU(4) state emerges in  
 capacitively coupled quantum dots, where it actually has been identified first.
In this device, shown in Fig.~\ref{fig:2dot}, the capacitively coupled dots 
can be described by the following simple Hamiltonian 
\be
H_{\rm dot} = {E_{C+}\over 2} (n_+ - N_{g+})^2 
+ {E_{C-}\over 2} (n_- - N_{g-})^2 + {\tilde E}_C n_+ n_-\;,
\ee
where the dimensionless gate voltages $N_{g\pm}$ set the number $n_+$ and $n_-$ of the 
electrons on the left and right dots, respectively. The last term is due to the 
capacitive coupling  between the two dots, and it is essentially this term which is 
 responsible for the SU(4) physics discussed.
As shown in Fig.~\ref{fig:2dot}.b, in the parameter space of the two-dot 
regions appear, where the two  states $(n_+,n_-)=(1,0)$ and  $(n_+,n_-)=(0,1)$ are 
almost degenerate, while the states $(n_+,n_-)=(0,0)$ and $(n_+,n_-)=(1,1)$ are pushed 
to higher energies of  order $\sim {\tilde E}_C$. 
In the simplest, however, most frequent case 
the states $(1,0)$ and  $(0,1)$ have both spin 
$S=1/2$, associated with the extra electron on the dots. 
Therefore, in the regime above for temperatures below the charging energy 
$\tilde E_C$  and the level spacing $\Delta$ of the dots, 
the dynamics of the double dot  is essentially restricted to the subspace 
$\{S^z = \pm 1/2; \; n_+-n_- = \pm 1\}$, and we can describe 
its charge fluctuations in terms of the {\em orbital pseudospin}  
$T^z \equiv (n_+ - n_-)/2 = \pm \frac12$.
 Coupling  the two dot system to leads, we arrive at the 
very same Hamiltonian as for the triangular dots, although with very different 
parameters.  Much of the previous discussions apply to this system
as well which, in addition to being a good spin-filter, 
also exhibits a giant magneto-resistance.\cite{su4} 
\begin{figure}[bht]
\centering
\includegraphics[width=5cm]{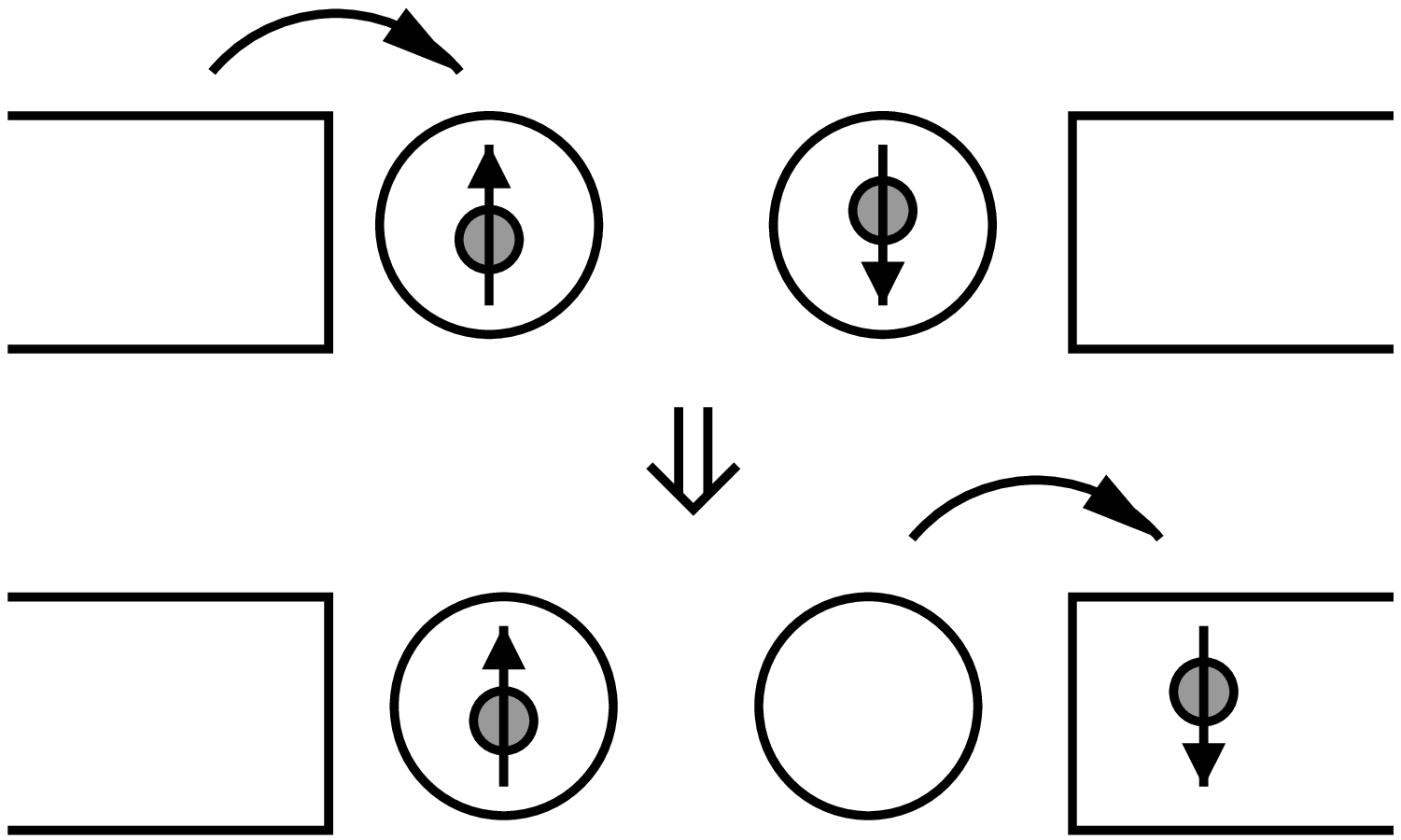}
\\
(a)
\\
\includegraphics[width=5cm]{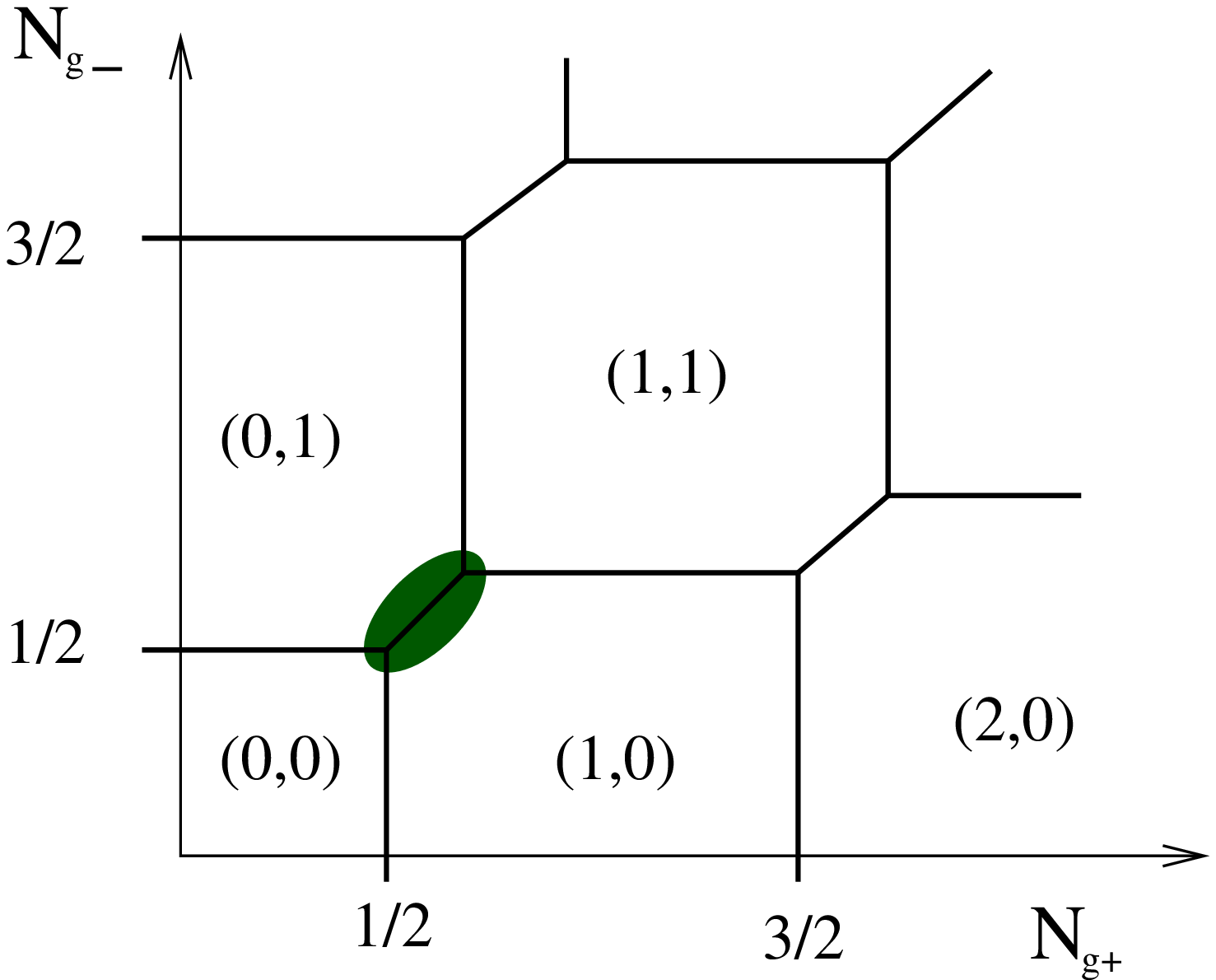}
\\
(b)
\caption{\label{fig:2dot} (a) A virtual process leading to entanglement 
between charge and spin fluctuations and the SU(4) Kondo state in the double 
dot device. 
(b) Charging states of the double dot device as a function of the dimensionless
gate voltages $N_{g\pm}$. The colored region indicates the regime where the two 
states $(1,0)$ and $(0,1)$ are almost degenerate.}
\end{figure}

\vskip0.2cm
{\bf
\parindent=0pt  Singlet-triplet transition
}

\vskip0.2cm

So far we discussed the case where there is a single conduction electron on the 
degenerate levels. The regime between the two middle peaks in Fig.~\ref{fig:Coulomb}
where there is two electrons on the 
(almost) degenerate multiplet is, however,  also extremely interesting.  

In this regime the Hund's rule coupling $J_H$ in Eq.~(\ref{eq:H_dot})
is very important. This coupling is typically smaller than the  
level spacing $\Delta$ in usual quantum dots. 
If, however, it is larger  than the separation 
between the last occupied and first empty levels, $\delta\epsilon$, then it 
gives rise to a triplet ground state with $S=1$. In particular, such a 
spin $S=1$ state forms if one puts two electrons on the 
orbitally degenerate level of a symmetrical quantum dot discussed before,\cite{triangle,Herrero}  
but almost  degenerate states may also occur in usual single electron 
transistors just by accident.\cite{vanderWiel}
Since in many cases one can shift the levels and thus tune $\delta \epsilon$ by an external 
magnetic field\cite{vanderWiel,Herrero} or simply by changing the shape of the dot 
by gate electrodes,\cite{KoganST} one can actually drive  
a quantum dot from  a triplet to a singlet state as illustrated in Fig.~\ref{fig:ST}.

\begin{figure}[htb] 
\centering
\includegraphics[width=5cm]{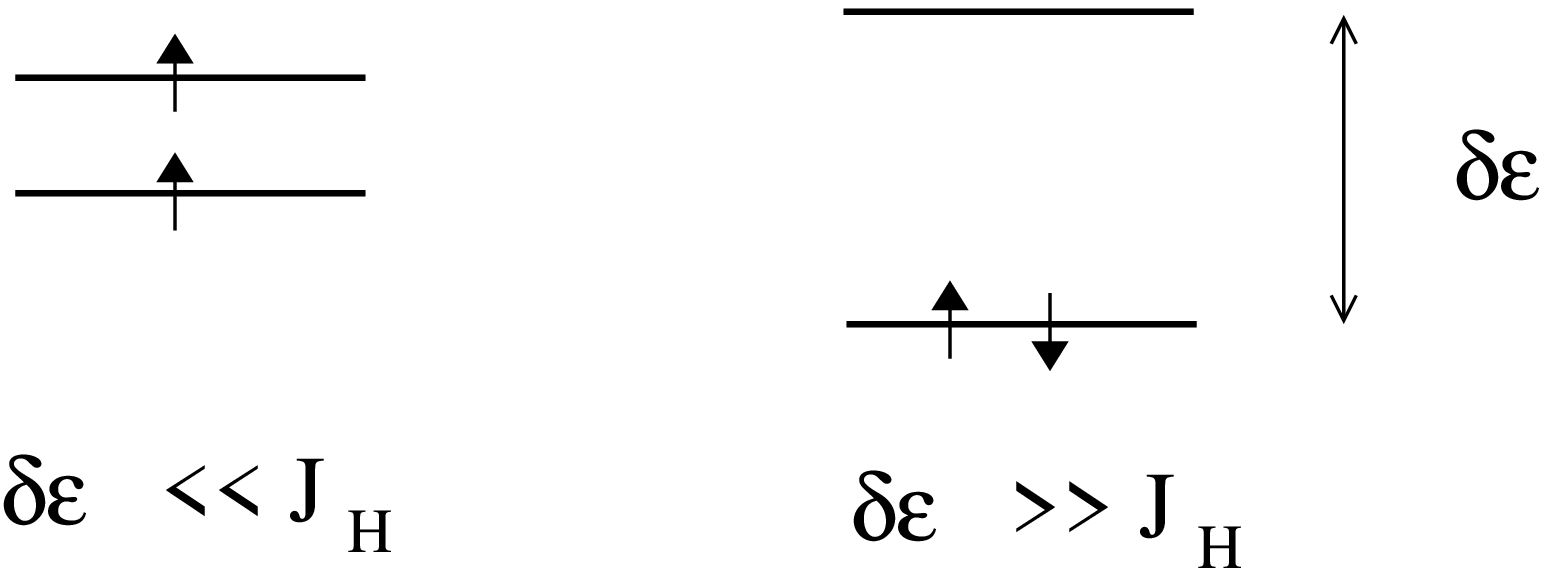}
\caption{\label{fig:ST} 
The state of a quantum dot changes from  a triplet to a singlet as the 
separation $\delta\epsilon$ between the last occupied and first empty level
increases.}
\end{figure}

While this transition has been first studied in  vertical dots,\cite{Sasaki,Pustilnik,Eto} 
here we shall focus on the  usual lateral arrangement of 
Fig.~\ref{fig:ST} which has very different transport 
properties.\cite{vanderWiel,real,VBH,HofstetterZarand} 
For the sake of simplicity let us assume that we have a completely 
symmetrical device and that the two levels $|\pm\rangle$ 
participating in the formation  of the triplet state are even and odd. 
In this case we can introduce the even and odd fields 
$\psi_\pm \equiv {(\psi_R \pm \psi_L)/ \sqrt{2}}$ by Eq.~(\ref{unitary}), 
which by symmetry only couple to the even and odd states, $|+\rangle$ and $|-\rangle$,
respectively. The hybridization term in this case simply reads
\bea
\hat V & = &  
 t_+ \sum_{\sigma} \int d\epsilon\; \Bigl( d_{+ \sigma}^\dagger
  \psi_{+ \sigma}(\epsilon)  + h.c.\Bigr)
\nonumber \\
&+&  t_- \sum_{\sigma} \int d\epsilon\; \Bigl( d_{- \sigma}^\dagger
  \psi_{- \sigma}(\epsilon)  + h.c.\Bigr)
\;.
\label{eq:V_ST}
\eea
To describe the isolated dot we can use the following simplified version of 
Eq.~(\ref{eq:H_dot}), 
\begin{eqnarray}
H_{\rm dot}  =  
\sum_{\sigma} \Bigl( \epsilon_+ d^\dagger_{+\sigma} d_{+ \sigma} +
\epsilon_- d^\dagger_{-\sigma} d_{- \sigma} \Bigr)
\nonumber  \\
-  {J_H\over 2} {\vec S}^{\;2}   +  {E_C \over 2} (n_+ + n_-)^2 \;.
\label{eq:H_dot_ST} 
\end{eqnarray}

It is instructive to study the triplet state of the dot first, 
$\delta \epsilon \ll J_H$. Second order perturbation theory 
in Eq.~(\ref{eq:V_ST})   in this regime gives the following Hamiltonian
replacing (\ref{eq:exchange})
\bea 
H_{\rm eff} & = & 
\sum_{\tau = \pm} \sum_{\sigma}  \sum_{\epsilon}\; \epsilon \;
\psi_{\tau, \sigma }^\dagger(\epsilon) \psi_{\tau,\sigma}(\epsilon) 
\label{eq:exchange_triplet}
\\
\nonumber
&+& {J_+\over 2} {\vec S} \; \sum_{\sigma,\sigma'}\int\int d\epsilon\;d\epsilon' 
\psi_{+\sigma}^\dagger(\epsilon) \vec \sigma_{\sigma\sigma'} 
\psi_{+\sigma'}(\epsilon') 
\\
\nonumber
&+& {J_-\over 2} {\vec S} \; \sum_{\sigma,\sigma'}\int\int d\epsilon\;d\epsilon' 
\psi_{-\sigma}^\dagger(\epsilon) \vec \sigma_{\sigma\sigma'} 
\psi_{-\sigma'}(\epsilon') \;.
\eea
Clearly, the even and odd electrons couple with  different exchange couplings $J_\pm$
to the spin.  However, now $\vec S$ is a spin $S=1$ operator, and to screen it  
completely, one needs to bind {\em two} conduction electrons to it. This implies that 
 an electron from both the even and the odd channels will be bound to the 
spin, and correspondingly two consecutive Kondo effects will take 
place at temperatures 
\be
T_{+} \approx \Delta\; e^{- 1/J_+}\;\gg \;T_{-} \approx \Delta\; e^{- 1/J_-}
\;.
\ee
This also implies that the conductance at $T=0$ temperature must 
vanish in the Kondo limit by the following simple argument:\cite{real}
Again, we can use the Friedel sum rule to obtain the $T=0$ temperature 
phase shifts $\delta_\pm=\pi/2$ in both the even and odd channels.
In the original left-right basis this implies that the lead electrons scatter 
as $\psi_{L/R} \to -\psi_{L/R}$, {\em i.e.}, their wave function vanishes at the dot 
position (by Pauli principle), and  $\psi_{L/R}$ are completely reflected. 

It is a simple matter to express the $T=0$ temperature conductance in 
terms of the $T=0$ temperature phase shifts by means of the Landauer-Buttiker 
formula as\cite{Buttiker}
\be
G = {e^2\over h}   \sum_\sigma \sin^2(\delta_{+,\sigma} - \delta_{-,\sigma} )\;.
\label{eq:conductance}
\ee
This formula immediately implies that the conductance as a function of a Zeeman field 
$B$ must be non-monotonic:\cite{real}
As we argued before, the conductance of the dot is small for $B=0$ at $T=0$ temperature.
For $T_-\ll B\ll T_+$, however, the Kondo effect in the odd channel is 
suppressed and correspondingly the phase shifts in this channel are approximately 
given by $\delta_{- \uparrow} \approx \pi$ and $\delta_{- \downarrow} \approx 0$, 
while the phase shifts in the even channel are still $\delta_{- \sigma}\approx \pi/2$,
and thus  by Eq.~(\ref{eq:conductance}) the conductance must be close to $2e^2/h$. 
For even larger magnetic fields,  $B\gg T_{+}, T_{-}$, the Kondo effect is killed in both 
channels, and correspondingly  $\delta_{\pm \uparrow} \approx \pi$ and 
$\delta_{\pm \downarrow} \approx 0$, resulting in a small conductance again.
The magnetic field dependence of the phase shifts obtained from a numerical renormalization 
group calculation and the corresponding conductance are shown 
in Fig.~\ref{fig:phase_shift+cond_B}.\cite{HofstetterZarand}  By general arguments,\cite{real}
 similar non-monotonic behavior must occur in the temperature- and bias-dependence
of the conductance, as it has indeed been observed experimentally.\cite{real,vanderWiel}

\begin{figure}[bht]
\begin{center}
\includegraphics[width=0.6\linewidth]{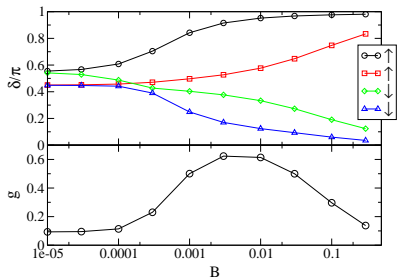}
\end{center}
\caption{
\label{fig:phase_shift+cond_B}
Phase shifts (top) and dip structure in the conductance  as a function of Zeeman field $B$ 
on the triplet side of the transition at zero temperature, computed by the 
numerical renormalizaion group method (from Ref.~\onlinecite{HofstetterZarand}). 
}
\end{figure}

Clearly, the $T=0$ temperature conductance must be also small on the singlet 
side of the transition,  $\delta\epsilon \gg J_H$, where the dot is in a singlet 
state and no Kondo effect occurs.  However, in the vicinity of the degeneracy 
point, $\delta\epsilon\approx J_H$,  the triplet and the singlet states of the 
dot are almost degenerate,  and quantum fluctuations between these four  
states  generate another type of
strongly correlated state with an increased Kondo temperature
and a large conductance.\cite{vanderWiel,HofstetterZarand,real}

To compute the full $T=0$ conductance as a function of $\delta \epsilon$
non-perturbative methods such as numerical renormalization 
group are needed.\cite{HofstetterZarand} In the vicinity of the transition 
point the conductance goes up to $2 e^2/h$, in perfect agreement with the 
experimental observations. While the non-monotonic behavior characteristic of 
the triplet state disappears in the vicinity of the transition, it reappears
on the singlet side. However, there the size of the dip is not determined 
by the smaller Kondo scale, $T_-$, but rather by the excitation 
energy of the triplet, $\sim \delta\epsilon - J_H$.\cite{vanderWiel,VBH,HofstetterZarand}
Note that the transition between the triplet and singlet states 
is smooth and the singlet-triplet transition is rather a cross-over than 
a phase transition in the above scenario.

\begin{figure}[htb]
\begin{center}
\includegraphics[width=0.6\linewidth]{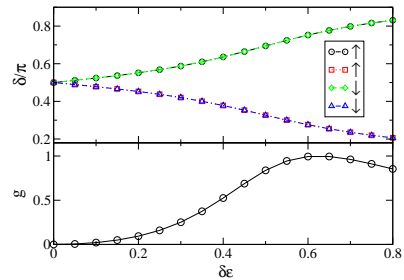}
\end{center}
\caption{
\label{fig:phase_shift+cond_Delta}
Phase shifts (top) and corresponding conductance (bottom) as a function of 
orbital splitting $\delta\epsilon$ at temperature $T=0$
(from Ref.~\onlinecite{HofstetterZarand}). 
}
\end{figure}


Also,  the  picture outlined above changes substantially 
if both states $|\pm\rangle$ happen  to have the same parity, and couple only to 
one of the fields $\psi_+$. In this case the conductance is small on the singlet side 
of the transition and exhibits a dip as a function of temperature/magnetic field.\cite{VBH} 
However, the spin of the dot cannot be screened 
on the triplet side even at $T=0$ temperature, and a real Kosterlitz-Thouless type
quantum phase transition occurs, where the $T=0$ temperature conductance has a jump at the 
transition point.\cite{VBH}  
The triplet phase in this case is also 
anomalous,\cite{VBH} and in fact is of a 'marginal Fermi liquid type',\cite{marginal,PiersPRL} 
since the spin is not fully screened.\cite{VBH,underscreened} 
Correspondingly, the conductance saturates very slowly,  and behaves 
asymptotically as $G\sim cst - 1/\ln^2(T_+/T)$. 
Similar behavior is expected to occur if the smaller Kondo temperature $T_-$ is much below  
the measurement temperature, and indeed a behavior in agreement with the Kosterlitz-Thouless 
scenario of  Ref.~\onlinecite{VBH} has  been observed in 
some experiments.\cite{KoganST} 


\section{Charge fluctuations and two-channel Kondo effect 
at the degeneracy point}
\label{Matveev}

In the previous section we focussed our attention to the regimes where charge 
fluctuations of the dot were negligible. In the vicinity of the degeneracy points, 
$N_g\approx\mbox{half-integer}$, however, this assumption is not valid, and charge 
fluctuations must be treated non-perturbatively. 

To have an insight how change fluctuations can lead to 
non-perturbative behavior, 
let us study the simplest circuit one can envision, 
the so-called single electron box (SEB), 
where only one lead is attached to a quantum dot (see Fig.~\ref{fig:SEB}).
Let us furthermore focus to the limit of small 
tunneling and $E_C\gg T \gg \Delta$.  The charging energy of the dot in this case 
is  given again by Eq.~(\ref{eq:H_C1}), and the tunnel coupling to the 
electrode  reads 
\be
\hat V = \sum_j \sum_\epsilon
\Bigl\{ t_j\; d^\dagger_{j \sigma } \psi_{\epsilon \sigma}
+ {\rm h.c.} \Bigr\}\;.
\label{tun}
\ee

\begin{figure}[tbh]
\includegraphics[width=8cm]{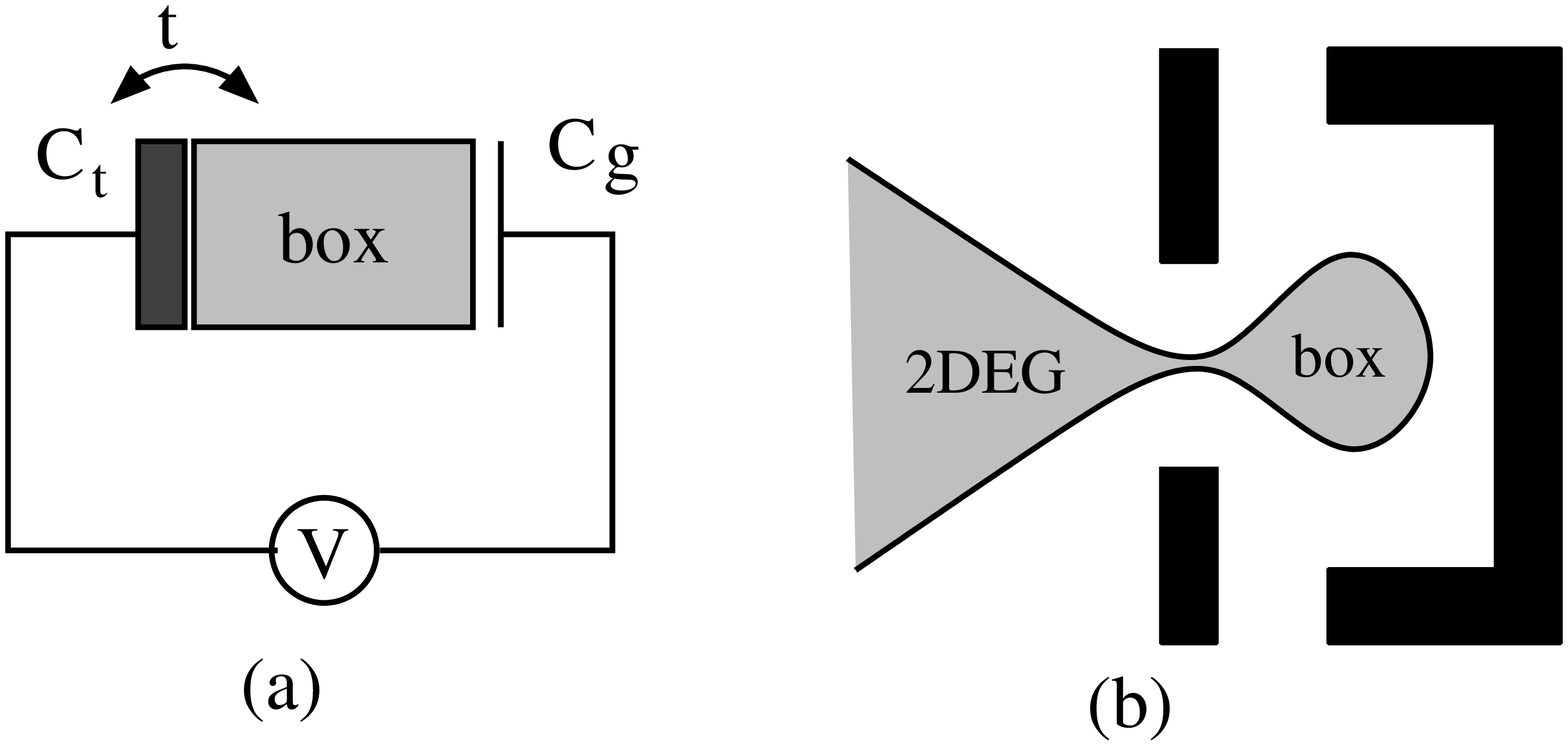}
\caption{
\label{fig:SEB} Sketch of the single electron box. Fig.~b. shows the top view of the 
 regime where electrons move in the the two-dimensional electron gas.  
Black areas indicate various  electrodes necessary to shape the electron gas. 
}
\end{figure}

Let us now focus our attention to 
the regime  $N_g \approx 1/2$, where with a good accuracy the charge 
on the dot fluctuates between $n_{\rm dot} = 0$ and $1$. 
In this regime tunneling processes generated by Eq.~(\ref{tun}) become 
correlated, because of the {\em constraint} that the charge of the dot must be either 
$0$ or $1$. To keep track of this constraint, we can introduce
the pseudospin operators,
 $T^+ \equiv |1\rangle\langle 0|$, 
$T^- \equiv |0\rangle\langle 1|$, and $T^z \equiv (|1\rangle\langle 1| - |0\rangle\langle 0|)/2$, 
and rewrite the tunneling part of the Hamiltonian and $H_{\rm dot}$ as 
\bea
\hat V  & = & \sum_j \sum_\epsilon
\Bigl\{ T^+ t_j\;   d^\dagger_{j \sigma } \psi_{\epsilon \sigma}
+ {\rm h.c.} \Bigr\}\;,
\label{tun2}
\\
H_{\rm dot} & = &-\Delta E \; T^z\;
\eea
where $\Delta E\sim (N_g - 1/2)$ is simply the energy difference between the two charging states 
of the dot. We can now attempt to compute the expectation value of the dot charge,
$\langle n_{\rm dot}\rangle$ in the regime $-1/2< N_g<1/2$ by doing 
perturbation theory in $\hat V$ to obtain  in the limit of vanishing level spacing,  
\be 
\langle n_{\rm dot}\rangle =  {g\over 4\pi^4}
 \ln\Bigl( {1 + 2 N_g\over 
1- 2N_g} \Bigr) + {\cal O}(g^2)\;,
\label{eq:n_dot}
\ee
with $g =  G/G_Q$ the dimensionless conductance of the junction. 
Although a finite level spacing 
cuts off the logarithmic singularity at $N_g = \pm 1/2$,
Eq.~(\ref{eq:n_dot}) clearly indicates that perturbation theory breaks 
down in the vicinity of the degeneracy points. 

In fact, following the mapping originally proposed by Matveev,\cite{Matveev}
we shall now show that the Hamiltonian above can be mapped to that of the two-channel Kondo 
problem. To perform the mapping, 
we rewrite the tunneling Hamiltonian in a more suggestive way 
by  introducing the new fields normalized by the density of states in the box
and in the lead, $\varrho_{\rm dot}$ and  $\varrho_{\rm lead}$, respectively
\be
D_{\sigma} \equiv {1\over \sqrt{\varrho_{\rm box} }} \sum_{j} d_{j,\sigma}\;,
\phantom{nn}
C_{\sigma} \equiv {1\over \sqrt{\varrho_{\rm lead} }} \sum_{\epsilon} \psi_{\epsilon,\sigma}\;,
\label{eq:fields}
\ee 
and organize them into a four component spinor
\be
\psi_{\tau,\sigma} \equiv \pmatrix{C_\sigma  \cr D_\sigma \cr}\;. 
\ee
In the limit $\Delta\to 0$ the tunneling amplitudes can be simply replaced by their 
average, $t_j \to  t\equiv \langle |t_j|^2\rangle_j^{1/2}$, and 
we can rewrite the tunneling part of Hamiltonian as
\be 
H_{\rm perp} =  {j_\perp \over 2} \sum_\sigma
  ( T^+ \psi^\dagger_{\sigma} \tau^- \psi_{\sigma} + {\rm h.c.})\;,
\label{eq:H_tun2}
\ee
where the operator $\tau^\pm$ just flips the orbital spin $\tau$ of the field $\psi_{\tau\sigma}$, and  
$j_\perp$  is a dimensionless coupling proportional to the tunneling, 
$j_\perp = 2 t \sqrt{\varrho_{\rm box} \varrho_{\rm lead}}$. Thus 
$j_\perp^2 \sim  t^2{\varrho_{\rm box} \varrho_{\rm lead}} $ is directly related to 
the dimensionless conductance $g$ of the tunnel junction
\be
g 
= {\pi^2 } j_\perp^2\;.
\ee

Eq.~(\ref{eq:H_tun2}) is just the 
Hamiltonian of an anisotropic two-channel Kondo model,\cite{Matveev,Cox}  
the orbital spins $T$ and $\tau$ 
playing the role of the spins of the original two-channel Kondo model, 
and the electron spin
$\sigma$ providing a silent channel index. 
The presence of this additional channel index (electron spin)
makes the physics of the two-channel Kondo model entirely different 
from that of the 
single channel Kondo problem, and leads to  non-Fermi liquid properties:\cite{Cox}
The low temperature conductance between the dot and the lead turns out to scale to 
$g(T\to0) \to e^2/h$ with a $\sqrt{T}$ singularity and 
the capacitance diverges logarithmically as $N_g\to 1/2$ and $T\to 0$.\cite{Matveev}
Thus  the theory of Matveev predicts that the sharp steps  
in $\langle n_{\rm dot}\rangle$ become smeared out as shown in Fig.~\ref{fig:steps}, 
although the slope of the steps diverges at the degeneracy points. 

This is, however, not the full picture. Very recently, Karyn le Hur studied the 
effect of dissipative coupling to other leads in the circuit, and showed
that if these additional leads are resistive enough, then the dissipation 
induced by them leads to a phase transition, where the steps are restored.\cite{LeHur} 
This transition can be shown by  bosonization methods to be of 
Kosterlitz-Thouless type.\cite{NoisyBox} In Fig.~\ref{fig:magnetization} we 
show the shape of the step computed using numerical renormalization group methods 
for the single  channel case  (spinless fermions) that clearly shows the above 
phase transition.\cite{NoisyBox}

\begin{figure}
\epsfxsize=6cm
\epsfbox{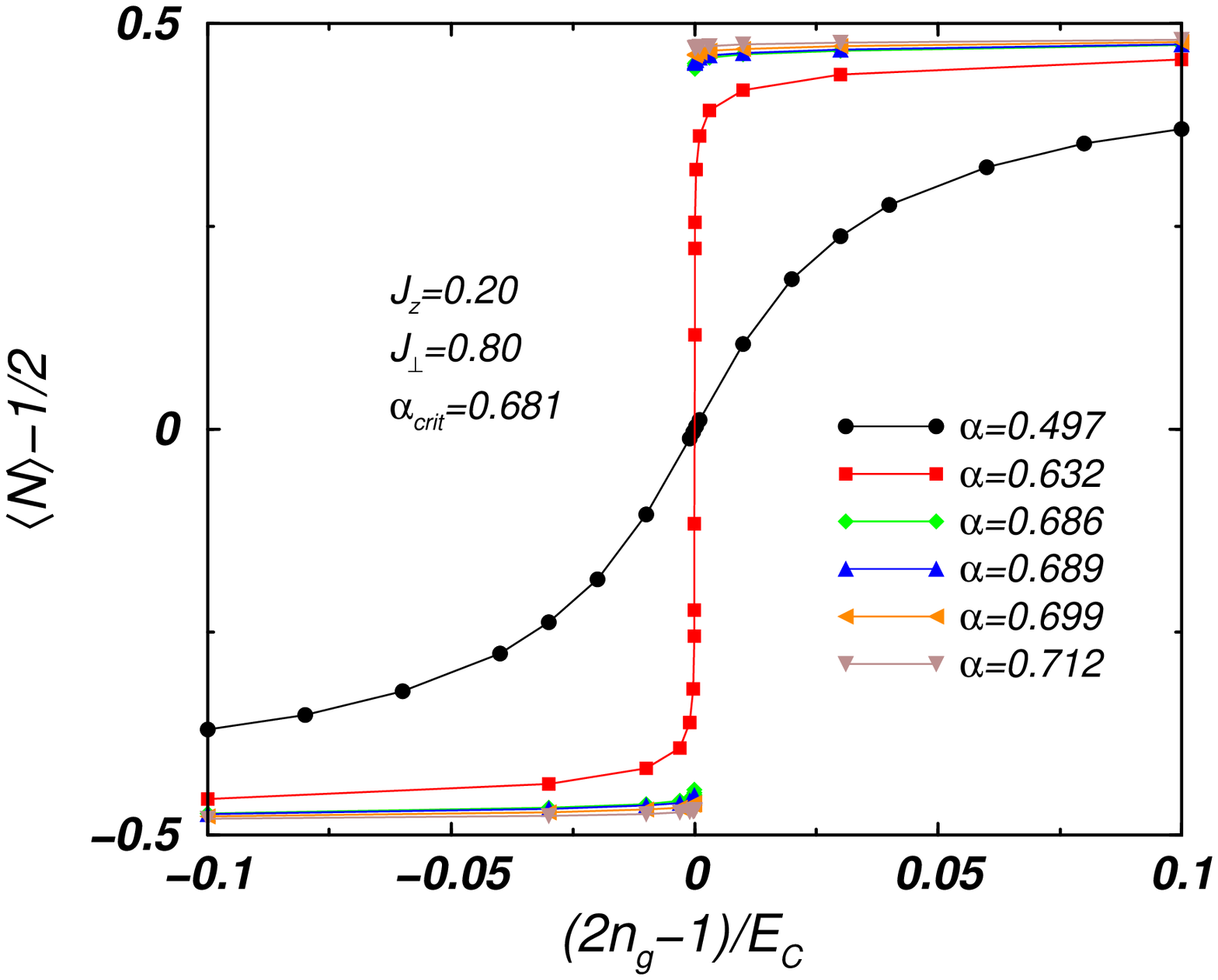}
\caption{
The expectation value of  $n_{\rm dot} - 1/2$ 
for  different values of $\alpha$. For moderate values of the bosonic coupling the step is
smeared out by quantum fluctuations of the charge of the dot. 
The step gets sharper and sharper  as $\alpha$ approaches  $\alpha_c$, and
For $\alpha > \alpha_c$ a jump appears in $\langle {\hat n}_{\rm dot} \rangle$
(from Ref.~\onlinecite{NoisyBox}).
} 
\label{fig:magnetization}
\end{figure}

We emphasize again, that the above  mapping holds only in the  
regime $\Delta< T,\omega,... < E_C$, where the level spacing of the box can 
be neglected, and is only valid for the case of a single mode contact,
Unfortunately, for small semiconductor dots with large enough $E_C$ 
the ratio $E_C/\Delta$ is not very large,\cite{Berman,Wilhelm}
and therefore  the regime where the two-channel Kondo behavior 
could be observed is rather limited. In fact, this intriguing
 two-channel Kondo behavior has never been observed convincingly 
experimentally.\cite{Berman} 

The ratio  $E_C/\Delta$ can be much larger in metallic 
grains,\cite{Devoret,Schoen} However, metallic grains have been 
connected so far only through leads with a large number of conduction 
modes, and the behavior of these systems is rather different
from what we discussed so far: 
They can be described by an infinite channel Kondo model,\cite{Schoen} 
which predicts, {\em e.g.},  that the junction conductance 
goes to zero logarithmically at the degeneracy point 
in the $T\to0$ limit,
$g_{\rm metal} (T)\sim 1/\ln^2(T/T_K)$, in clear disagreement 
with the two-channel Kondo predictions.  It turns out, that 
for a {\em finite} number $N_c$ of conductance modes a new 
temperature scale $T^*$ appears:\cite{Wilhelm} Above this scale the conductance
 decreases as $\sim 1/\ln^2(T)$, while below this scale 
it increases and approaches the two-channel Kondo value 
$g = e^2/h$. Unfortunately, this scale goes to zero exponentially   
fast with the number of modes, $T^* \sim  \exp(- C\;N_c)$. Therefore, one really needs
to prepare single mode contacts to observe the two-channel Kondo 
behavior in metallic grains, which is a major experimental challenge.

Recently, another realization of the two-channel Kondo behavior 
has also been proposed in semiconductor devices, where one has a 
somewhat better control of the Kondo scale and  $E_C/\Delta$ than in 
single electron boxes.\cite{Oreg,Eran} 

So far we discussed only the limit of small level spacing $\Delta\to 0$.
The physics of the degeneracy point $N_g=1/2$ 
remains also rather non-trivial in the limit $T\ll \Delta$.  
In this regime the dot behaves as a mixed valence 
atom, and charge fluctuations are huge.\cite{Hewson} 
The description of these mixed valence fluctuations is a rather 
complicated problem, and requires non-perturbative methods
such as the application of 
numerical renormalization group.\cite{Theo}

\section{Conclusions}

In the present paper we reviewed some of the 
interesting strongly correlated states that can be realized using 
quantum dots. These artificial structures behave in many respects
like artificial atoms, excepting the difference in the energy 
scales that characterize them. 
Having a full control over these devices
opened up the possibility 
of building structures that realize 
 unusual strongly correlated states like the ones discussed
in this paper, that are very difficult to observe in atomic physics.
This technology  also enabled one 
to study   out of equilibrium and transport properties
of such individual 'atoms'. 

However, building  quantum dots and quantum wires 
is just the first step towards a new technology, 
which  aims to  construct devices from {\em real} atoms and 
molecules instead of mesoscopic structures. Although 
a real breakthrough took place in recent years, and molecules 
have been contacted and used to construct single 
electron transistors,\cite{molecular_SET} the 
technology is far  from being controlled.

Building (gated) quantum dot arrays in a controlled 
way has not been solved satisfactorily yet either.
Having a handle on such systems  could give way to study 
 quantum phase transitions
in lattices of artificial atoms in the laboratory.
Experimentalists are also facing the challenge of
producing  {\em hybrid structures} at the nanoscale: 
these structures can hopefully be used in  future spintronics 
applications or quantum computing. 

There is a lot of open questions 
on the theoretical side as well: As we mentioned already, 
the problem of treating strongly correlated systems 
in  out of equilibrium is unsolved 
even for simple toy models, and it is still a dream to study 
molecular transport through correlated out of equilibrium 
atomic clusters by combining  {\em ab initio} and many-body 
methods. Moreover,  many important questions like the 
interplay of ferromagnetism and  strong correlations in ferromagnetic grains
have not been studied in sufficient detail.

I would like to thank all my collaborators, especially 
Laszl\'o Borda, Walter Hofstetter, and A. Zawadowski
for the valuable discussions.  
This research has been supported by
Hungarian Grants No. OTKA T038162, T046267, and T046303, 
and the European 'Spintronics' RTN HPRN-CT-2002-00302.

\appendix
\vspace{0.5cm}
\section{Hartree approximation for a quantum dot}
\label{app:Hartree}

To derive Eq.~(\ref{eq:H_C1}), let us consider 
a metallic grain within the Hartree approximation.   Then the wave functions 
$\varphi_j$ must be determined self-consistently by solving 
the following equations:
\bea
\Bigl( -{\Delta\over 2m} + V_H(\vec r) \Bigr) 
\varphi_j = E_j \; \varphi_j\;,
\nonumber
\\
V_H(\vec r)  =  V(\vec r) + \int d{\vec r}' U({\vec r} - {\vec r}') \varrho({\vec r}')\;,
\label{eq:V_H}
\eea
where $V$ is the confining potential generated by the positively charged ions, 
$U$ is the electron-electron interaction, and 
$V_H$ is the Hartree potential. The electronic density $ \varrho({\vec r})$ in 
Eq.~(\ref{eq:V_H}) must be computed  as $ \varrho({\vec r}) = \sum_{j,\sigma} n_{j,\sigma} \; 
|\varphi_j|^2$, with $n_{j,\sigma}$ the occupation number of the levels. 

The total energy of the system with $N =  \sum_{j,\sigma} n_{j,\sigma} $ electrons is
\be
E^N_H = \sum_{j,\sigma} E_j\; n_{j,\sigma} -{1\over 2} 
 \int d{\vec r}\; d{\vec r}' \varrho({\vec r}) 
V({\vec r} - {\vec r}') \varrho({\vec r}')\;,
\ee
where the second term compensates for overcounting the electron-electron interaction.
In order to compute the energy cost of adding another electron to the grain, 
we should solve selfconsistently 
Eqs.~(\ref{eq:V_H}) for $ \sum_{j,\sigma} n_{j,\sigma} = N+1 $, compute 
$E^{N+1}_H$ and then determine the difference between these two energies,
$E^{N+1}_H- E^{N}_H$. However, one can approximately compute this energy 
by just noticing that  the charge of the extra electron in 
 state $N+1$  must go to the {\em surface} 
of the grain to produce an approximately constant potential inside the grain,
and is screened within a layer of the Fermi wavelength $\sim \lambda_F$.  
The change in  the Hartree potential is simply given as 
\be 
\delta V_H   =  \int d{\vec r}' U({\vec r} - {\vec r}') \;\delta\varrho({\vec r}')\;.
\ee
But since  the change of the electronic density can be very well approximated 
by a classical {\em surface charge} for a grain size $L\gg \lambda_F$, 
  $\delta V_H $ is just the classical potential of the charged grain.
Consequently, $\delta V_H \approx e^2/C$ inside the metallic grain. 
Using this simple fact
we find that 
adding an extra electron to the grain 
shifts all Hartree energies as $E_j\to E_j + e^2/C$ and  requires  an energy 
\be 
E^+ \approx E_f + {e^2\over 2 C}\;,
\ee
with $ E_f $ the Hartree energy of the first unoccupied level, and 
$C$ the classical capacitance of the grain. Defining the chemical potential 
as the Hartree energy of the last occupied level, $\mu \equiv \epsilon_l$ 
and defining the quasiparticle energies  as $\epsilon_j \equiv E_j - \mu$ 
we find by extending the above analysis to the excited states as well that 
the energy of the dot is approximately described by 
Eqs.~(\ref{eq:dot1}) and (\ref{eq:H_C1}).
Note that for an isolated dot this analysis gives 
$N_g = - \mu C/e^2$, which is usually not equal to zero, so adding and removing an electron
requires different energies, just as in case of an atom.

\end{document}